\documentclass[12pt]{article}

\usepackage{latexsym}
\newcommand{\beqn}{\begin{eqnarray}}
 \newcommand{\eeqn}{\end{eqnarray}}

 \newcommand{\be}{\begin{equation}}
 \newcommand{\ee}{\end{equation}}
 \newcommand{\ba}{\begin{array}}
 \newcommand{\ea}{\end{array}}
 
 \newcommand{\pa}{\partial}
 \newcommand{\re}{\ref}
 \newcommand{\ci}{\cite}
 \newcommand{\ds}{\displaystyle}
 \newcommand{\la}{\label}
 \newcommand{\bfr}{\begin{flushright}}
 \newcommand{\efr}{\end{flushright}}
 
 \newcommand{\rIm}{{\rm Im\5}}
 \newcommand{\rRe}{{\rm Re\5}}
\newcommand{\bfl}{\begin{flushleft}}
\newcommand{\efl}{\end{flushleft}}
\newcommand{\fr}{\frac}

\newcommand{\ov}{\overline}

\newcommand{\ti}{\tilde}

\newcommand{\loota}{\hbox{\enspace{\vrule height 7pt depth 0pt width
      7pt}}}
\newcommand{\bo}{{\hfill\loota}}

\renewcommand{\Pr}{{\bf Proof~}}

\newcommand{\bB}{{\bf B}}
\newcommand{\bC}{{\bf C}}

\newcommand{\bP}{{\bf P}}
\newcommand{\bR}{{\bf R}}

\newcommand{\bF}{{\bf F}}

\newcommand{\M}{{\cal M}}
\newcommand{\cO}{{\cal O}}

\newcommand{\cS}{{\cal S}}

\newcommand{\ve}{\varepsilon}

\newcommand{\de}{\delta}

\newcommand{\al}{\alpha}
\newcommand{\ga}{\gamma}

\newcommand{\si}{\sigma}
\newcommand{\om}{\omega}
\newcommand{\Om}{\Omega}
\newcommand{\na}{\nabla}

\newcommand{\lam}{\lambda}
\newcommand{\Lam}{\Lambda}
\newcommand{\ka}{\kappa}

\newcommand{\bmuo}{\mbox{\boldmath $\mu_{\omega}$}}
\newcommand{\bmuone}{\mbox{\boldmath $\mu_{\omega_1}$}}

\newcommand{\5}{{\hspace{0.5mm}}}

\newcommand{\ddd}{\stackrel{.\kern-.07em.\kern-.07em.}}
\newcommand{\bre}{|\kern-.25em|\kern-.25em|}
\def\R{{\rm I\kern-.1567em R}}                              
 \def\C{{\rm C\kern-5.9pt                                    
 \vrule height 7.5pt width 0.4pt depth -0.8pt \phantom {.}}}
 \def\Z{{\sf Z\kern-4.5pt Z}}                                

 \renewcommand{\theequation}{\thesection.\arabic{equation}}
\renewcommand{\thesection}{\arabic{section}}
\renewcommand{\thesubsection}{\arabic{section}.\arabic{subsection}}
\newtheorem{theorem}{Theorem}[section]
\newtheorem{qtheorem}{QTheorem}[section]

\renewcommand{\thetheorem}{\arabic{section}.\arabic{theorem}}
\newtheorem{defin}[theorem]{Definition}

\newtheorem{lemma}[theorem]{Lemma}
\newtheorem{example}[theorem]{Example}
\newtheorem{examples}[theorem]{Examples}
\newtheorem{exercice}[theorem]{Exercise}
\newtheorem{remark}[theorem]{Remark}
\newtheorem{remarks}[theorem]{Remarks}
\newtheorem{cor}[theorem]{Corollary}
\newtheorem{pro}[theorem]{Proposition}

\newcommand{\bd}{\begin{defin}}
 \newcommand{\ed}{\end{defin}}
\newcommand{\bt}{\begin{theorem}}
 \newcommand{\et}{\end{theorem}}
\newcommand{\bqt}{\begin{qtheorem}}
 \newcommand{\eqt}{\end{qtheorem}}

\newcommand{\bp}{\begin{pro}}
 \newcommand{\ep}{\end{pro}}

\newcommand{\bl}{\begin{lemma}}
 \newcommand{\el}{\end{lemma}}
\newcommand{\bc}{\begin{cor}}
 \newcommand{\ec}{\end{cor}}

\newcommand{\bex}{\begin{example}}
 \newcommand{\eex}{\end{example}}
\newcommand{\bexs}{\begin{examples}}
 \newcommand{\eexs}{\end{examples}}

\newcommand{\bexe}{\begin{exercice}}
 \newcommand{\eexe}{\end{exercice}}

\newcommand{\br}{\begin{remark} }
 \newcommand{\er}{\end{remark}}
\newcommand{\brs}{\begin{remarks}}
 \newcommand{\ers}{\end{remarks}}

\newcommand{\const}{\mathop{\rm const}\nolimits}

\newcommand{\mod}{\mathop{\rm mod}\nolimits}

\begin{document}


  \begin{titlepage}

\bigskip\bigskip\bigskip

\begin{center}
 {\Large\bf
On Asymptotic Stability of Solitary Waves
\bigskip\\
in a Nonlinear Schr\"odinger Equation}
\vspace{1cm}
\\
{\large V.~S.~Buslaev}\footnote{
Supported partly by RFBR research grants
 05-0101076 and 05-01002944.
}
\\
{\it
Department of Mathematical and
Computational Physics
\\
Faculty of Physics\\
St-Petersburg University, Russia
}\\
e-mail:~buslaev@mph.phys.spbu.ru
\bigskip\\
{\large A.~I.~Komech}
\footnote{
 On leave Department of Mechanics and Mathematics,
Moscow State University, Moscow 119899, Russia.
Supported partly
by FWF grant P19138-N13 and Max-Planck 
Institute for Mathematics in the Sciences (Leipzig).}\\
{\it Fakult\"at f\"ur Mathematik, Universit\"at Wien\\
Nordbergstrasse 15, Wien A-1090, Austria}\\
 e-mail:~alexander.komech@univie.ac.at
\medskip\\
{\large E.~A.~Kopylova}
\footnote{Supported partly by  FWF grant P19138-N13 and
      RFBR grant 06-01-00096.}\\
{\it Keldysh Institute for Applied Mathematics RAS\\
Miusskaya sq.4.,Moscow 125047, Russia}\\
  e-mail:~ek@vpti.vladimir.ru
\medskip\\
{\large D.~Stuart}
\footnote{Partially supported by EPSRC grant A00133/01}\\
{\it Centre for Mathematical Sciences,\\
Wilberforce Road, Cambridge, CB3 OWA}\\
 e-mail:~D.M.A.Stuart@damtp.cam.ac.uk
\end{center}

\date{}

\markboth{V.Buslaev, A.Komech, E.Kopylova}
  {On Asymptotic Stability of Solitary Waves}

\vspace{0.5cm}
\begin{abstract}
The long-time asymptotics is analyzed for finite energy solutions of the
1D  Schr\"odinger equation  coupled to a nonlinear oscillator.
The coupled system is invariant with respect to the phase
rotation group $U(1)$.
For initial states close to a solitary wave, the solution converges to a sum
of another solitary wave and dispersive wave which is a solution to the free
Schr\"odinger equation.
The proofs use the strategy of Buslaev-Perelman \ci{BP1,BP2}:
the linerization of the dynamics on the solitary manifold,
the symplectic orthogonal projection and method of majorants.
\end{abstract}

\end{titlepage}
\setcounter{equation}{0}
\section{Introduction}
\label{intr}

Our main goal is the study of the distinguished dynamical role of the "quantum stationary states"
for a model $U(1)$-invariant nonlinear Schr\"odinger equation
\be\la{S}
  i\dot\psi(x,t)=
  -\psi''(x,t)-\de(x)F(\psi(0,t)),\quad x\in\R.
\ee
Here $\psi(x,t)$ is a continuous complex-valued wave function
and $F$ is a continuous function, the dots stand for the derivatives in $t$
and the primes in $x$.
All derivatives and the equation are understood in the distribution sense.
Physically, equation  (\re{S}) describes the system of the free Schr\"odinger equation
coupled to an oscillator attached at the point $x=0$:
$F$ is a nonlinear ``oscillator force''.

We assume that $F(\psi)=-\na U(\psi)$ where $U(\psi)=u(|\psi|)$. Then
(\re{S}) defines a $U(1)$-invariant Hamilton system
and admits finite energy  solutions of type
$\psi_\om(x)e^{i\om t}$ called  {\it solitary waves} or {\it nonlinear eigenfunctions}.
The  solitary waves constitute a two-dimensional {\it solitary manifold} in the Hilbert
phase space of finite energy states of the system. We prove the asymptotics of type
\be\la{sol-as-i}
     \psi(\cdot,t)\sim\psi_{\om_{\pm}}e^{i\om_{\pm}t}+W(t)\Phi_{\pm},
     \quad t\to\pm\infty,
\ee
where $W(t)$ is the dynamical group of the free Schr\"odinger equation,
$\Phi_{\pm}\in C_b(\R)\cap L^2(\R)$ are the corresponding asymptotic scattering states,
and the remainder converges to zero as $\cO(|t|^{-1/2})$ in global norm of
$C_b(\R)\cap L^2(\R)$. Here $C_b(\R)$ is the space of bounded continuous functions
$\R\to \C$. The asymptotics hold for the solutions with initial states close to the
{\it stable part} of the solitary manifold, extending the results of
\ci{BP1,BP2,MW96,PW92,PW94} to the equation (\re{S}).

Let us note that we impose conditions which are more general
than the standard ones in the following respects:
\smallskip\\
i)
We do not hypothesize any spectral properties of the linearized equation,
and do not require any smallness condition on the initial state (only
closeness to the solitary manifold).
\\
ii) The stable part of the solitary manifold is characterized by a condition
on the nonlinearity (\ref{stab}).
The relation of this to the standard criterion for
orbital stability $\ds\pa_\om \!\!\int \!|\psi_\om(x)|^2dx >0$
(see \ci{GSS} and references therein) will be discussed below.
\smallskip\\
This progress is possible on account of the simplicity of our model
which allows an exact analysis of all spectral properties of the linearization.

Our investigation is inspired by a fundamental problem of quantum mechanics.
The solitary waves were introduced by Schr\"odinger for the quantum electron
coupled to the Maxwell field,  \ci{Sch}. He identified the solitary waves with
the quantum stationary states. The asymptotics of type (\re{sol-as-i}) for the coupled
$U(1)$-invariant Maxwell-Schr\"odinger equations would describe the transitions
between the quantum stationary states, while the dispersive wave $W(t)\Phi_{\pm}$ would
correspond to the electromagnetic radiation. Let us note however, that
the asymptotics of type (\re{sol-as-i}) are not proved yet for the coupled equations.

For the first time, the asymptotics of type (\re{sol-as-i}) were established
by Soffer and Weinstein \ci{SW1,SW2} (see also \ci{PW}) for nonlinear  $U(1)$-invariant
Schr\"odinger equation with small initial states if the nonlinear coupling constant is
sufficiently small.
The next result was obtained by Buslaev and Perelman \ci{BP1, BP2}
who proved that the solitary manifold attracts finite energy solutions
of a 1D nonlinear  $U(1)$-invariant translation invariant Schr\"odinger equation
with initial states sufficiently close to the {\it stable part} of the solitary manifold.

The novel techniques of Buslaev and Perelman are based on the symplectic geometry in
Hilbert space and the spectral theory of nonselfadjoint operators.
These techniques were motivated by the investigation of soliton asymtotics for
integrable equations (a survey can be found in \ci{DIZ} and \ci{FT}),
and by the methods introduced by Soffer and Weinstein \ci{SW1,SW2,W85}.
Similar techniques were developed by Miller, Pego and
Weinstein for the 1D modified KdV and RLW equations, \ci{MW96,PW92,PW94}.
The methods and results were extended in \ci{BP3, BS} to the
Schr\"odinger equations with more complicated spectral properties, and
in \ci{IKV05} to a translation invariant wave-particle system.
Further references can be found in \ci{BS} and \ci{IKV05}.

Let us comment on the general  strategy of our proofs.
We develop the approach \ci{BP2,IKV05} for our problem. Firstly, we apply the
{\it symplectic projection} onto the solitary manifold to separate the motion along
the solitary manifold and in transversal direction. Secondly, we derive the
{\it modulation equations} for the parameters of the symplectic
projection, and linearize the transversal dynamics at the projection of
the trajectory. The linearized equation is nonautonomous, and this is one of the
fundamental difficulties in the proof. This difficulty is handled by the introduction
of an autonomous  equation (by freezing the time) with an application of the modulation
equations to estimate the resulting additional error terms. A principal role in the rest
of the proof is played by the uniform decay of the frozen linearized dynamics projected
onto the continuous spectrum, and the method of majorants.

Let us note the following two main novelties in our approach to the uniform decay.
First, we calculate exactly all needed spectral properties
of corresponding generator.
Second, we do not use a spectral representation of the generator.
Instead, we develop the Jensen-Kato approach applying directly the
Zygmund type Lemma 6.1 (cf. \ci[Lemma 10.2]{JK})  to the Laplace integral of the resolvent.
We expect that the development would be promising for more general problems.

The paper is organized as follows. In Section \ref{ndsec} some notation
and definitions are given. In Section \ref{swsec} we describe all nonzero solitary waves
and formulate the main theorem. The linearization at a solitary wave is
carried out in Section \ref{linsec}. In Sections \ref{laplace} and \ref{subspace},
we construct the spectral representation for the linearized equation.
In Section \ref{rpsec} we establish the time decay for the linearized
equation in the continuous spectrum.
In Section \ref{modsec} the modulation equations for the parameters of
the soliton are displayed. The decay of the transverse component is proved in
Sections \ref{sassec},\ref{prsec}.
In Section \ref{solas-sec} we obtain the soliton asymptotics (\ref{sol-as-i}).
In Appendix we study the resolvent of linearized equation.

In conclusion, we expect that the asymptotics (\ref{sol-as-i}) holds
for {\it any} finite energy solution of the equation (\ref{S}), however this
is still open problem.
We hope to prove it combining our methods  with the techniques of the papers
\ci{KoKo06,KK06}, where {\it global attraction} to the solitary manifold is proved
for the 1D Klein-Gordon equation with the same delta-nonlinearity. We
also intend to treat in a later publication the case when the
linearization has nontrivial {\it stable} oscillatory modes (which occurs if
(\ref{stab}) holds but (\ref{Co}) does not).

\setcounter{equation}{0}
\section{Notation and definitions}
\label{ndsec}
We identify a complex number $\psi=\psi_1+i\psi_2$ with the real two-dimensional
vector $(\psi_1,\psi_2)\in\R^2$ and assume that the vector version
$\bf F$ of the oscillator
force $F$ admits a real-valued potential,
\be\la{P}
  {\bf F}(\psi)=-\na U(\psi),\quad\psi\in\R^2,  ~~~~U\in C^2(\R^2).
\ee
Then (\re{S}) is formally a Hamiltonian system with Hamiltonian
\be\la{H}
 {\cal H}(\psi)=\fr 12
 \int  |\psi'|^2 dx+U(\psi(0)).
\ee
which is conserved for sufficiently regular finite energy solutions.
We assume that the potential $U(\psi)$ satisfies the inequality
\be\la{U}
   U(z)\ge A-B|z|^2 \quad {\rm with\; some}\quad A\in\R,\quad B>0.
\ee
Our key assumption concerns the $U(1)$-invariance
of the oscillator  (cf \ci{BL}),
where
$U(1)$ stands for the
group $e^{i\theta}$, $\theta\in[0,2\pi]$ acting by phase rotation
$\psi\mapsto e^{i\theta}\psi$.
Namely, we assume that
$U(\psi)=u(|\psi|^2)$ with $u\in C^2(\R)$.
Therefore,
by (\re{P}),
\be\la{I}
   F(\psi)= a(|\psi|^2)\psi,\quad\psi\in\C\5,~~~~~~~~a\in C^1(\R),
\ee
where $a(|\psi|^2)$ is real.
Then
$F(e^{i\theta}\psi)=e^{i\theta} F(\psi),\quad\theta\in[0,2\pi] $ and $F(0)=0$
for continuous $F$. Obviously, the symmetry holds true if
$ U(\psi)=u(|\psi|^2)$.
The symmetry implies that $e^{i\theta}\psi(x,t)$ is a solution to
(\re{S})  if $\psi(x,t)$ is. The equation is $U(1)$-invariant in
the sense of \ci{GSS}, and the N\"other theorem implies the {\it charge} conservation:
\be\la{Q}
  {\cal Q}(\psi)=\int |\psi|^2 dx=\const.
\ee
The main subject of this paper is an analysis of the special role
played by
``quantum stationary states'', or {\it solitary waves} in the sense of \ci{GSS},
which are  finite energy solutions of the form
\be\la{SW}
  \psi(x,t)=\psi_\om(x)e^{i\om t},\quad \om\in\R.
\ee
The frequency  $\om$ and the amplitude $\psi_\om(x)$ solve the following
{\it nonlinear eigenvalue problem}:
\be\la{NEP}
  -\om\psi_\om(x)= -\psi_\om''(x)-\de(x)F(\psi_\om(0)),\quad x\in\R.
\ee
which follows directly from  (\re{S}) and (\re{I})
since $\om\in\R$.

\bd\la{dSS}
  $\cS$ denotes the set  of all nonzero solutions
  $\psi_\om(x)\in H^1(\R)$ to (\re{NEP}) with all possible $\om\in\R$.
\ed
Here $H^1(\R)=H^1$ denotes the Sobolev space of complex valued measurable functions
with $\ds\int (|\psi'|^2+|\psi|^2\bigr)dx<\infty$.
We give below in section \ref{swsec} a complete analysis
of the set $\cS$ of all nonzero solitary waves $\psi_\om(x)$ by an explicit calculation:
it consists of functions $C(\om)e^{-\sqrt\om |x|+i\theta}$
with $C>0,\,\om=\om(C)>0$
and any $\theta\in [0,2\pi]$, and $C$ restricted to lie in a set which, in the case of
polynomial $F$, is a finite union of one-dimensional intervals.
Notice that $C=0$ corresponds to the zero function $\psi(x)=0$ which
is always a solitary  wave  as $F(0)=0$,
and for $\om\le 0$ only the zero solitary wave exists.


Our main results describe the large time behavior of the global solutions
whose existence is guaranteed by the following theorem, which is proved in \cite{KoK06}.
\begin{theorem}\label{locex}
   i)
   Let conditions (\re{P}), (\re{U}) and  (\re{I}) hold. Then
   for any $\psi_0(x)\in H^1$ there exist a unique solution
   $\psi(x,t)\in C_b(\R,H^1)$
   to the equation (\ref{S}) with initial condition $\psi(x,0)=\psi_0(x)$.\\
ii)
There exists $\Lambda(\psi_0)>0$
such that
the following a priori bound holds:
\be
\sup\limits\sb{t\in\R}\Vert{\psi(t)}\Vert\sb{H^1}
\le \Lam(\psi_0)<\infty.
\label{a-priori-h1}
\ee
\end{theorem}


The functional spaces we are going to consider are the weighted Banach spaces
$L^p_{\beta}$, $p\in [1,\infty)$, $\beta\in\R$ of complex valued
  measurable functions with the norm
\be\la{norm}
  \Vert u\Vert_{L^p_{\beta}}=\Vert (1+|x|)^{\beta} u(x)\Vert_{L^p}.
\ee


\setcounter{equation}{0}
\section{Solitary waves and statement of the main theorem}
\label{swsec}
\bl
The set of all nonzero solitary waves is given by
$$
\cS=\Bigl\{\psi_{\om}e^{i\theta}=C e^{i\theta-\sqrt\om |x|}:
\;
\om>0,~~~C>0,~~~
\sqrt\om =a(C^2)/2>0,~~~~\theta\in[0,2\pi]\Bigr\}.
$$
\el
\Pr
Let us calculate  all solitary waves (\re{SW}).
The equation (\re{NEP}) implies $\psi''(x)=\om\psi(x)$, $ x\ne 0$,
hence the formula $\psi(x)=C_\pm e^{\sqrt{\om} x}$ gives two linearly independent
solutions in each of the two regions $\pm x>0$ depending on which branch of
$\sqrt\om$ is chosen.
Since $\psi(x)\in L^2$ it is necessary that $\om> 0$  and the branch is chosen
with $\pm\sqrt\om>0$ for $\pm x<0$. Furthermore,
since  $\psi'(x)\in L^2$, the function $\psi(x)$ is continuous, hence
$C_-=C_+=C$ and the solutions are of the form
\be\la{swa}
  \psi(x)=C e^{-\ka |x|},\quad \ka=\sqrt{\om}> 0, \quad\om>0.
\ee
Finally we get an algebraic equation for the constant $C$  equating the coefficients
 of $\de(x)$ in both sides of (\re{NEP}):
\be\la{ais}
  0=\psi'(0+)-\psi'(0-)+F(\psi(0)).
\ee
This implies $0=-2\ka C+F(C)$, or equivalently,
\be\la{swac}
  \ka=\fr{F(C)}{2C}=\fr {a(C^2)} 2.
\ee
\bo
\bc
The set $\cS$ is a smooth manifold with co-ordinates $\theta\in\R\mod 2\pi$
and $C>0$ such that $a(C^2)>0$.
\ec
\br\la{romega}
We will analyse only the solitary waves with $a'(C)\ne 0$.
On the manifold $\cS$
we have $\om=\kappa^2$ with $\kappa=a(C^2)/2$
according to (\re{swac}).
Hence,
the parameters $\theta, \om$
locally also are smooth coordinates
on $\cS$ at the points with $a'=a'(C)\ne 0$ since
$\om'=2\kappa\kappa'=aa'C\ne 0$ then.
\er
The soliton solution  is a trajectory
$\psi_{\om(t)}(x)e^{i\theta(t)}=Ce^{-\sqrt{\om(t)} |x|}e^{i\theta(t)}$,
where the parameters  satisfy the equation $\dot\theta=\om$, $\dot\om=0$.
The solitary waves $e^{i\theta}\psi_\om(x)$ map out in time an orbit
of the $U(1)$ symmetry group. This group acts on the phase space $H^1({\bf R})$ preserving
${\cal H}$ and the standard symplectic form (\ref{symp}); in other words the solitary waves
(\ref{SW}) are relative equilibria of the corresponding Hamiltonian system.

Let us denote  $N(C)=\ds\int|\psi_\om(x)|^2dx$ with $\om=\ka^2$, and
$\kappa=a(C^2)/2$ according to (\re{swac}).
It is easy to compute that $N(C)=C^2/\kappa$. We now differentiate:
$$
  N'(C)=\fr{2C}\kappa-\fr{C^2\kappa'}{\kappa^2}.
$$
Differentiating the identity (\re{swac}), we obtain $\kappa'=a'C$. Thus, again by (\re{swac}),
$$
  N'(C)=\fr{2C}{\kappa}(1-\fr{a'C^2}a)\not=0
$$
if $C> 0$, $a>0$ and $a'\not=a/C^2$. Therefore noticing that
$N'(C)=\om'(C)\partial_\om\ds\int|\psi_\om|^2dx$ with $\om'(C)=2\kappa\kappa'=aa'C$,
we obtain the following result
\begin{lemma}\la{int-dif}
For $C> 0$, $a>0$ we have
$$
  \pa_\om\int|\psi_\om(x)|^2 dx< 0\quad\hbox{if}\;\;a'\in (-\infty,0)\cup (a/C^2,+\infty),
  \la{syc}
$$
and
$$
  \pa_\om\int|\psi_\om(x)|^2 dx>0\quad\hbox{if}\;\; 0<a'<a/C^2.\la{syc2}
$$
\end{lemma}
\br
{\rm
(i) Orbital stability of solitary waves is a much studied subject
(see \cite{GSS} for very general theorems in this area, and \cite{St01} for an
approach more similar to that taken in this paper).
The standard condition for orbital stability (\cite{GSS}) for the
present problem would read $\ds\pa_\om \int |\psi_\om(x)|^2dx
>0$; this is expected to be a necessary and sufficient condition
for orbital stability when the Hessian of the
augmented Hamiltonian (\cite{St01}) has a single negative
eigenvalue. In the present problem it can be easily calculated that
this Hessian is non-negative when $a'<0$ and thus the standard
condition is not necessarily relevant if $a'<0$. Indeed Theorem
\ref{main} asserts stability in the case $a'<0$.  Restricting to
$a'>0$, in which case the Hessian does have a single negative
eigenvalue, the calculation above shows that orbital stability is
expected to hold when $a'<a/C^2$. In this paper we will work under the
spectral condition (\ref{Co}) which, for $a'>0$, is slightly stricter:
it is imposed to ensure that the linearization has no discrete
spectrum except zero (which is always present on account of the
circular symmetry of the problem).  If $a/\sqrt 2 C^2<a'<a/C^2$ there are
two purely imaginary eigenvalues of the linearized operator. It is
intended to treat this case in a later publication thus extending our
proof of asymptotic stability to the entire range
\be
  -\infty<a'<a/C^2.\label{stab}
\ee
For $a'>a/C^2$ the linearized operator has a
positive eigenvalue and the solitary wave is linearly unstable.

(ii) It is explained at the end of section \ref{linsec} that (\ref{syc})
can be interpreted as saying the restriction of the standard
symplectic form (\ref{symp}) to the tangent space to $\cS$ is
non-degenerate (i.e. $\cS$ satisfies the condition to be a symplectic
submanifold).
}
\er
\bd\la{spcon}
  We say the solitary wave $\psi_{\om}(x)e^{i\theta}=Ce^{-\sqrt{\om} |x|+i\theta}$,
  $C> 0$ satisfies the spectral condition if $\om>0$ and (cf. Remark \re{romega})
  \be\la{Co}
     a'(C^2)\in (-\infty,0)\cup(0,a(C^2)/(\sqrt 2C^2)).
  \ee
\ed
Let us denote by $W(t)$ the dynamical group of the free Schr\"odinger equation:
$W(t)f$ is defined by the Fourier representation for all tempered distributions $f$.
Our main  theorem is the following:
\begin{theorem}\label{main}
   Let conditions (\re{P}), (\re{U})  and  (\re{I}) hold,
   $\beta\ge 2$ and $\psi(x,t)\in C(\R,H^1)$ be the solution to the equation (\ref{S})
   with initial value $\psi_0(x)=\psi(x,0)\in H^1\cap L^1_\beta$ which
   is close to a solitary wave
   $\psi_{\om_0}e^{i\theta_0}=C_0e^{-\sqrt{\om_0}|x|+i\theta_0}$
   with  $C_0>0$ and $\om_0>0$:
   \be\la{close}
     d:=\Vert\psi_0-\psi_{\om_0}e^{i\theta_0}\Vert_{H^1\cap L^1_{\beta}}\ll 1.
   \ee
   Assume further that the spectral condition (\ref{Co}) holds for the solitary wave with $C=C_0$.
   Then for $d$ sufficiently small the solution admits the following asymptotics:
   \be\la{sol-as}
     \psi(\cdot,t) = \psi_{\om_{\pm}}e^{i\om_{\pm}t}+W(t)\Phi_{\pm}
     +r_{\pm}(t),\quad t\to\pm\infty,
   \ee
   where $\Phi_{\pm}\in C_b(\R)\cap L^2(\R)$ are the corresponding asymptotic
   scattering states, and
   \be\la{rate}
   \Vert r_{\pm}(t)\Vert_{C_b(\R)\cap L^2(\R)}=\cO(|t|^{-1/2}),\quad t\to\pm\infty.
   \ee
\end{theorem}
\br
It is possible to derive further information about the structure of
$\Phi_\pm$ and $r_\pm(t)$  as discussed towards the end of section 10.
\er


\setcounter{equation}{0}
\section{Linearization on the solitary wave}
\la{linsec}

As the first step in the proof of main theorem, let us linearize the nonlinear
Schr\"odinger equation (\re{S}) at a solitary wave $e^{i(\om t+\theta)}\psi_\om(x)$, with
$ \psi_\om(x)=Ce^{-\ka|x|}$ where $\ka=\sqrt{\om}> 0$ and $C> 0$.
Substituting
\be\la{prol}
  \psi(x,t)=e^{i(\om t+\theta)}(\psi_\om(x)+\chi(x,t))
\ee
to (\re{S}), we obtain,
\be\la{lin1}
  -\om\chi(x,t)+i\dot\chi(x,t)=-\chi''(x,t)-\de(x)[F(C+\chi(0,t))-F(C)]
\ee
Use the representation (\re{I}) to write
\beqn\la{lin2}
  F(C+\chi)-F(C)&=&a(|C+\chi|^2)(C+\chi)-a(|C|^2)C\nonumber\\\nonumber\\
  &=&a((C+\chi)(\ov C+\ov\chi))(C+\chi)-a(|C|^2)C\nonumber\\\nonumber\\
  &=&a(|C|^2)\chi+a'(|C|^2)C
  (C\ov\chi+\ov C\chi)+\cO(|\chi|^2)\nonumber\\\nonumber\\
  &=&a(C^2)\chi+a'(C^2)C^2(\ov\chi+\chi)+\cO(|\chi|^2)
\eeqn
since $C\ge 0$.
Hence, the first order part of (\re{lin1}) is given by
\beqn\la{lin3}
   i\dot\chi(x,t)&=&-\chi''(x,t)+\om\chi(x,t)\nonumber\\\nonumber\\
   &&-\de(x)[a(C^2)\chi(0,t)+a'(C^2)C^2 2\rRe\chi(0,t)].
\eeqn
Now it is evident that the first order part is not linear over the complex field.
On the other hand, it is linear over the real field.
Hence, it would be useful to rewrite (\re{lin3}) in the real form.
Namely, identify $\chi=\chi_1+i\chi_2\in\C$  with the real vector
$(\chi_1,\chi_2)\in\R^2$ and denote it again by $\chi$.
Then (\re{lin3}) becomes
\beqn\la{lin3r}
  j\dot\chi(x,t)&=&-\chi''(x,t)+\om\chi(x,t)\nonumber\\\nonumber\\
  &&-\de(x)[a(C^2)+2a'(C^2)C^2 P_1]\chi(0,t),
\eeqn
where $P_1$ is the projector in $\R^2$ acting as
$\left(\ba{l}\chi_1\\ \chi_2\ea\right)\mapsto \left(\ba{l}\chi_1\\ 0\ea\right)$
and $j$ is the $2\times 2$ matrix
\be\la{j}
   j=\left(
   \ba{rr}
   0  &  -1\\
   1  &   0
   \ea
   \right)~~~~~~~~~~~~~~~~~~~~~~~~~~~~~~~~
\ee
Respectively, we also rewrite (\re{S}) in the  real form
\be\la{SV}
  j\dot\psi(x,t)= -\psi''(x,t)-\de(x)\bF(\psi(0,t)),
\ee
as an equation for $\psi(x,t)\in\R^2$
with
$\bF(\psi)\in\R^2$ which is the real vector version of $F(\psi)\in\C$.
Then the linearization (\re{lin3r}) reads as
\be\la{SVl}
  j\dot\chi(x,t)= -\chi''(x,t)+\om\chi(x,t)
  -\de(x)\bF'((C,0))\chi(0,t).
\ee
Here $\bF'$ is the differential of the map $\bF:\R^2\to\R^2$,
\be\la{aa1}
 \bF'((C,0))=
a+bP_1,~~~~~~~~~~ a:=a(C^2),~~~b:=2a'(C^2)C^2.
\ee

In order to apply the Laplace transform the next step is to complexify
the system (\re{SVl}) i.e. to consider it as a system of equations for the
complex functions $\chi_1(x,t),\chi_2(x,t)$, so $\chi(x,t)\in\C \5\5^2$ for any
fixed $(x,t)$. This gives a system which is linear over  the
complex field allowing application of the Laplace transform.
To write this system more concisely
let us denote the complex linear operator
\be\la{B}
   {\bf B}=-\ds\fr{d^2}{dx^2}+\om-\de(x)\bF'((C,0))=
   \left(
   \ba{cc}
   {\bf D}_1  &       0    \\
        0     &   {\bf D}_2
   \ea
   \right),
\ee
where
\be\la{D}
   \ba{l}
   {\bf D}_1=-\ds\fr{d^2}{dx^2}+\om-\de(x)[a+b],\\ \\
   {\bf D}_2=-\ds\fr{d^2}{dx^2}+\om-\de(x)a.
   \ea
\ee
The system (\re{SVl}) then reads as
\be\la{lin4}
   \dot\chi(x,t)={\bf C}\chi(x,t),~~~~~{\bf C}:=j^{-1}\bB=
   \left(
   \ba{rr}
    0          &    {\bf D}_2\\
   -{\bf D}_1  &        0
   \ea
   \right).
\ee
Theorem \ref{locex} generalises to the equation (\ref{lin4}):
the equation admits unique solution $\chi(x,t)\in C_b(\R,H^1)$
for every initial function $\chi(x,0)=\chi_0\in H^1$.
Denote by $e^{{\bf C}t}$ the dynamical group of equation (\ref{lin4})
acting in the space $H^1$.


\setcounter{equation}{0}
\section{Laplace transform}
\label{laplace}
Equation (\ref{lin4}) can be solved by the Laplace transform
$\ti\chi(x,\om):=\ds\int_0^\infty e^{-\lam t}\chi(x,t)dt$. The Laplace transform
is analytic function in the complex halfplane $\rRe\lam>0$
with the values in $H^1$
since the solution is bounded in $H^1$.
This implies that the resolvent $\bR(\lam):=(\bC-\lam)^{-1}$ is also analytic
for  $\rRe\lam>0$, with values in the space of bounded operators on $H^1$.
From the inversion of the Laplace transform  we obtain
\be\la{LT}
   e^{{\bf C}t}=-\frac 1{2\pi i}\int\limits_{-i\infty}^{i\infty}
    e^{\lam t}{\bf R}(\lam+\ve)~d\lam,~~~~~~~t>0,
\ee
for any $\ve>0$,
where the integral converges in the sense of distributions of $t\in\R$.

We assume that the spectral condition (\ref{Co}) holds from now on.
Then the resolvent admits analytic continuation from  $\rRe\lam>0$ to
the complex plain with the cuts
${\cal C}_+=[i\om,i\infty)$, ${\cal C}_-=(-i\infty,-i\om]$,
and with the pole of order two at $\lam=0$
as detailed in Appendix A.
Furthermore,
for $\lam\in{\cal C}_{+}\cup{\cal C}_{-}$,
 the resolvent ${\bf R}(\lam\pm\ve)$
has right and left limits ${\bf R}(\lambda\pm 0)$ as $\ve\to 0$.
Then (\ref{LT}) implies that
\be\la{IL}
   e^{{\bf C}t}=-\frac 1{2\pi i}\int\limits_{|\lam|=r}
   e^{\lam t}{\bf R}(\lam)~d\lam-
   \frac{1}{2\pi i}\int\limits_{{\cal C}_+\cup{\cal C}_-}
   e^{\lam t}({\bf R}\bigl(\lam+0)-{\bf R}(\lam-0)\bigr)~d\lam,\quad
   {\rm for~~any}\quad r\in(0,\omega)
\ee
by the Cauchy theorem.
Setting $t=0$, we obtain that
\be\la{ILg}
  1=-\frac 1{2\pi i}\int\limits_{|\lam|=r}
   {\bf R}(\lam)~d\lam-
   \frac{1}{2\pi i}\int\limits_{{\cal C}_+\cup{\cal C}_-}
   ({\bf R}\bigl(\lam+0)-{\bf R}(\lam-0)\bigr)~d\lam=\bP^0+\bP^c,
\ee
where
$\bP^0$
and $\bP^c$ stands for the corresponding
Riesz projectors
(see \ci{RN})
onto, respectively, the generalised null space of $\bC$ , and onto the
 continuous spectral subspace.
We will show in the next section that $\bP^0$  is the symplectic projection,
and therefore,  $\bP^c$ is also the symplectic
projection.
The projectors ${\bf P}^0$, ${\bf P}^c$ commute
with $\bC$ and with the group ${e^{{\bf C}t}}$. Let us note that
\be\la{eP0}
 \bP^0e^{{\bf C}t}=-\frac 1{2\pi i}\int\limits_{|\lam|=r}
  e^{\lam t} {\bf R}(\lam)~d\lam,~~~~
  {\bf P}^c e^{{\bf C}t}
  =-\frac 1{2\pi i}\int\limits_{{\cal C}_+\cup{\cal C}_-}
  e^{\lam t} ({\bf R}\bigl(\lam+0)-{\bf R}(\lam-0)\bigr)~d\lam.
\ee
The first
equation holds since
both sides are one-parameter groups of operators ,
and their derivatives at $t=0$ coincide.
The second equation follows from  (\ref{IL}) and the fact that $1=\bP^0+\bP^c$
by (\ref{ILg}).
Therefore, (\ref{IL}) becomes
\be\la{ILb}
   e^{{\bf C}t}= \bP^0e^{{\bf C}t}+\bP^ce^{{\bf C}t}.
\ee


\setcounter{equation}{0}
\section{Invariant subspace of discrete spectrum}
\label{subspace}
Here we prove that  $\bP^0$  is the symplectic projection onto the tangent space
of the solitary manifold  $\cS$ at the solitary wave $e^{j\theta}\psi_\om$.
The real form of the solitary wave is $e^{j\theta}\Phi_\om$
where $\Phi_\om=(\psi_{\omega}(x),0)$.
The tangent space to $\cS$ at the point $e^{j\theta}\Phi_\om$ with parameters
$\omega,\theta$ is the linear span of the derivatives with respect to $\theta$ and
$\omega$ cf. Remark \re{romega}) i.e.
$$
   T_{\omega,\theta}\cS\equiv\hbox{linear span}\Bigl\{je^{j\theta}\Phi_{\omega}(x),
   e^{j\theta}\partial_\omega\Phi_{\omega}(x)\Bigr\}.
$$
Notice that the operator $\bC$ corresponds to $\theta=0$ since we have extracted the
phase factors $e^{i\theta}$ from the solution in the process of linearization
(\re{prol}). The tangent space to $\cS$ at the point $\Phi_\om$
with parameters $(\omega,0)$ is spanned by the vectors
\be\la{deft}
  T_0(\om):=j\Phi_\om,\quad T_1(\om):=\partial_\omega\Phi_\omega.
\ee
Observe that
(\ref{NEP}) and its derivative in $\om$ give the following identities:
\be\la{diffeq}
  {\bf D}_2\psi_\omega=0\qquad {\bf D}_1(\partial_\omega\psi_\omega)=-\psi_\omega.
\ee
These formulae imply that the vectors $T_0$ and $T_1$ lie in the generalised
null space of the non-self-adjoint operator ${\bf C}$ defined in
(\ref{lin4}) and in fact Theorem \ref{Spectr} $ii)$ implies:
\begin{lemma}\label{gk}
   Let the spectral condition (\ref{Co}) hold.
   Then the generalised null space of ${\bf C}$ is two dimensional,
   is spanned by $T_0,T_1$, and
   $$
   {\bf C} T_0=0\qquad {\bf C} T_1=T_0.
   $$
\end{lemma}

We also introduce the symplectic form $\Omega$ for the real vectors
$\psi$ and $\eta$ by the integral
\be\la{symp}
   \Om(\psi,\eta)=\ds\int\langle j\psi,\eta\rangle dx
   =\ds\int\bigl(\psi_1\eta_2-\psi_2\eta_1\bigr) dx,
\ee
where $\langle\cdot,\cdot,\rangle$ stands for the scalar product in $\R^2$.
Since  $a'\not=a/C^2$  then by Lemma \ref{int-dif}
\be\label{nondeg}
   \bmuo=-\Omega(T_0,T_1)=\frac{1}{2}\partial_\om\int |\psi_\om|^2dx\not=0.
\ee
Hence,
the symplectic form $\Omega$ is nondegenerate on
the tangent space
$T_{\om,0}{\cal S}$, i.e. $T_{\om,0}{\cal S}$ is a symplectic subspace.
Therefore, there exists
a symplectic projection operator from $L^2({\R})$ onto $T_{\om,0}{\cal S}$.
\begin{lemma}\la{P0-def}
The operator ${\bf P}^0$, defined in (\ref{ILg}), is precisely
the symplectic projector from $L^2({\R})$ onto $T_{\om,0}{\cal S}$,
and, furthermore, it may be represented by the formula
\be\label{defsp}
  {\bf P}^0\psi=b_0 T_0+b_1 T_1\quad\hbox{with}\quad
  -\bmuo b_0=\Omega(\psi, T_1),\quad \bmuo b_1=\Omega(\psi, T_0).
\ee
\end{lemma}
{\bf Proof}
The coincidence of both definition (\ref{ILg}) and (\ref{defsp}) of operator
${\bf P}^0$ follows by the  Cauchy residue theorem from the formulas (\ref{nR})-(\ref{tP})
for the resolvent.
\hfill\bo.
\bc
$\bP^c=1-\bP^0$ is also symplectic projector.
\ec
\begin{remark}
  Since $T_0(\om), T_1(\om)$ lie in $H^1({\R})$ the operator ${\bf
  P}^0$ extends uniquely
  to define a continuous linear map $H^{-1}({\R})\to
T_{\om,0}{\cal S}$, which is still
  designated ${\bf P}^0$. In particular this operator can be applied
  to the Dirac measure $\delta(x)$.
\end{remark}

Using the Taylor expansion for the $e^{\lam t}$ at $\lam=0$
and the identity $\lam{\bf R}(\lam)={\bf CR}(\lam)-1$, we obtain by (\ref{eP0})
\be\la{P0-rep}
   {\bf P}^0e^{{\bf C}t}=(1+\bC t){\bf P}^0
\ee
\br
On the generalised null space itself ${\bf C}^2=0$ by Lemma \ref{gk} and so the
semigroup $e^{t{\bf C}}$ reduces to $1+{\bf C}t$ as usual for the exponential of the
nilpotent part of an operator.
\er

\setcounter{equation}{0}
\section{Time decay in continuous spectrum}
\label{rpsec}
From formulas (\ref{ILb}, (\ref{P0-rep}) we see that the solutions
$\chi(t)= e^{{\bf C}t}\chi_0$ of the linearized equation (\ref{lin4})
do not decay as $t\to\infty$ if $\bP^0\chi_0\ne 0$.
On the other hand, we do expect time decay of ${\bf P}^c\chi(t)$,
as a consequence of the Laplace representation (\ref{eP0}) for
${\bf P}^c e^{{\bf C}t}$:
\be\la{Pc-rep}
 {\bf P}^c e^{{\bf C}t}
  =-\frac 1{2\pi i}\int\limits_{{\cal C}_+\cup{\cal C}_-}
  e^{\lam t} ({\bf R}\bigl(\lam+0)-{\bf R}(\lam-0)\bigr)~d\lam.
\ee
The decay for the oscillatory integral is obtained from the analytic
properties of $\bR(\lam)$ for $\lam\in{\cal C}_+\cup{\cal C}_-$.
The resolvent $\bR(\lam)$ is an  integral operator
with matrix valued integral kernel
\be\la{nR}
  {\bf R}(\lam,x,y)=\Gamma(\lam,x,y)+P(\lam,x,y),
\ee
where the columns of matrices $\Gamma$ and $P$ are given in (\ref{GI}), (\ref{PI}),
(\ref{GII}), (\ref{PII}):
\be\la{tG}
   \Gamma(\lam,x,y)=\left( \ba{cc}
   \ds\frac 1{4k_+}  &  -\ds\frac 1{4k_-}\\
   \ds\frac i{4k_+}  &  \ds\frac i{4k_-}
   \ea\right)\left( \ba{cc}
   e^{ik_+|x-y|}-e^{ik_+(|x|+|y|)}  &  -i(e^{ik_+|x-y|}-e^{ik_+(|x|+|y|)})\\\\
   e^{ik_-|x-y|}-e^{ik_-(|x|+|y|)}  &   i(e^{ik_-|x-y|}-e^{ik_-(|x|+|y|)})
   \ea\right),
\ee
\be\la{tP}
   P(\lam,x,y)=\frac 1{2D} \left( \ba{cc}
   e^{ik_+|x|}    &   e^{ik_-|x|}\\
   ie^{ik_+|x|}   &  -ie^{ik_-|x|}
   \ea\right)
   \left( \ba{cc}
   i\al-2k_-       &    i\beta\\
   -i\beta         &   -i\al+2k_+
   \ea\right)
   \left( \ba{cc}
   e^{ik_+|y|}      &  -ie^{ik_+|y|}      \\
   e^{ik_-|y|}       &  ie^{ik_-|y|}
   \ea\right).
\ee
Here $k_\pm(\lam)=\sqrt{-\om\mp i\lam}$ is
the square root defined with cuts in the complex  $\lam$ plane
so that $k_\pm(\lambda)$ is analytic on ${\C}\setminus {\cal C}_\pm$ and
${\rm Im}\5 k_\pm(\lambda)>0 $ for $\lam\in\C\setminus {\cal C}_\pm$.
The constants $\al$, $\beta$ and $D=D(\lam)$ are given by the formulas
$$
 \al=a+b/2,\; \beta=b/2,\; D=2i\al(k_++k_-)-4k_+k_-+\al^2-\beta^2.
$$
Recall from Section \ref{polres} that $D(\lam)\ne 0$ for $\lam\in{\cal C}_+\cup{\cal C}_-$.
Clearly in order to understand the decay of ${\bf P}^c e^{t{\bf C}}$,
it is crucial to study the behaviour of $ {\bf R}(\lam,x,y)$ near the
branch points $\lam=\pm i\om$ (where $k_\pm$ vanish).

We deduce time decay for the group ${\bf P}^c e^{t{\bf C}}$
by means of the following  version of Lemma 10.2 from \cite{JK},
which is itself based on Zygmund's lemma \ci[p.45]{Z}.

Let ${\cal F}:[0,\infty)\to {\bf B}$ be a $C^2$ function with values in a
Banach space ${\bf B}$.
Let us define the ${\bf B}$-valued function
$$
I(t)=\int\limits^{\infty}_0 e^{-it\zeta}{\cal F}(\zeta)~d\zeta.
$$
\begin{lemma}\la{K}
 Suppose that ${\cal F}(0)=0$, and
for some $\delta>0$
  \be\la{Z1}
    {\cal F}''\in L^1(\delta,\infty; {\bf B}),
  \ee
  and
  \be\la{Z2}
    {\cal F} ''(\zeta)={\cal O}(\zeta^{p-2}),\quad \zeta\downarrow 0
  \ee
  in the norm of $~{\bf B}$ for some $p\in (0,1)$.
  Then $I(t)\in C_b(\ve,\infty;\bB)$ for any $\ve>0$, and
  $$
  I(t) ={\cal O}(t^{-1-p})
  \quad{\rm as}\quad t\to\infty
  $$
  in the norm of    $~~{\bf B}$.
\end{lemma}
For $\beta\ge 2$ let us introduce a Banach space ${\cal M}_\beta$, which is the subset
of distributions which are  linear combinations of $L^1_\beta$ functions and
multiples of the Dirac distribution at the origin with the norm:
\be\la{cM}
  \Vert \psi+C\delta(x)\Vert_{{\cal M}_\beta}:=\Vert\psi\Vert_{L^1_\beta}+|C|.
\ee
We will apply Lemma \ref{K} to the function
${\cal F}(\lam)={\bf R}(\lam+0)-{\bf R}(\lam-0)$ with values in the Banach space
${\cal B} =B({\cal M}_\beta,L^{\infty}_{-\beta})$ ,
the space of continuous linear maps ${\cal M}_\beta\to
L^\infty_{-\beta}$ for any $\beta\ge 2$.
\begin{theorem}\la{TD}
  Assume that the spectral condition (\ref{Co}) holds so that
  $\lambda=0$ is the only point in the discrete spectrum of the
  operator $\bC=\bC(\om)$. Then
for $\beta\ge 2$
  \be\la{decayR}
    \Vert {\bf P}^c e^{{\bf C}t}\Vert_{\cal B}
    ={\cal O}(t^{-3/2}),\quad t\to\infty.
  \ee
\end{theorem}
First we
use the formulas (\ref{Pc-rep}) and (\ref{nR}) to obtain
\be\la{Z}
  -2\pi i{\bf P}^c e^{{\bf C}t}=
  \int\limits_{{\cal C}_+\cup{\cal C}_-}\!
  e^{\lam t}(\Gamma(\lam+0)-\Gamma(\lam-0))\5d \lam
 ~~ +\int\limits_{{\cal C}_+\cup{\cal C}_-}\!
  e^{\lam t}(P(\lam+0)-P(\lam-0))\5 d\lam
\ee
Next we apply Lemma \ref{K} to each summand in the RHS of (\ref{Z})
separately. Then Theorem \ref{TD} immediately follows from the  two lemmas below.
\begin{lemma}\la{decayG}
If the assumption of Theorem \ref{TD} hold then
  \be
    \int\limits_{{\cal C}_+\cup{\cal C}_-}
    e^{\lam t}(\Gamma(\lam+0)-\Gamma(\lam-0))~d\lam
    ={\cal O}(t^{-3/2}),\quad t\to\infty
  \ee
  in the norm $\cal B$.
\end{lemma}
{\bf Proof }
We consider only the integral over ${\cal C}_+$ since the integral over ${\cal C}_-$
can be handled in the same way.
The point $\lam=i\om$ is the branch point for $k_+$, therefore,
if $\lam\in{\cal C}_+$ then since $k_-$ is continuous across ${\cal C}_+$
$$\Gamma(\lam+0)-\Gamma(\lam-0)=\Gamma^+(\lam+0)-\Gamma^+(\lam-0),$$
where $\Gamma^+$ is the  sum of those terms in $\Gamma$ which involve $k_+$.
Let us consider, for example,  $\Gamma^+_{11}$.
The expression (\re{tG}) implies for $y>0$ that
$$
  \Gamma^+_{11}(\lam,x,y)= \left\{ \ba{ll}
  0,  \!&\!x\le0,\\\\
  \ds\frac{e^{ik_+ y}(e^{-ik_+ x}-e^{ik_+ x})}{4k_+},  \!&\!0\le x\le y,\\\\
  \ds\frac{e^{ik_+ x}(e^{-ik_+ y}-e^{ik_+ y})}{4k_+},  \!&\!x\ge y.
  \ea \right.
$$
For $\lam\in {\cal C}_+$, the root
$k_+=\sqrt{-\om-i\lam}$ is real, and $k_+(\lam+0)=-k_+(\lam-0)$.
Then, for $y>0$,
\beqn
  \Gamma^+_{11}(\lam+0,x,y)-\Gamma^+_{11}(\lam-0,x,y)&=&
  -\Theta(x)\ds\frac{\sin(\sqrt{\zeta}|x|)\sin(\sqrt{\zeta}|y|)}{\sqrt{\zeta}},
\eeqn
where $\zeta=-\om-i\lam$, and $\Theta(x)=1$ for $x>0$ and zero otherwise.
The second derivative of the function
$f(\zeta)=\ds\frac{\sin(\sqrt{\zeta}|x|)\sin(\sqrt{\zeta}|y|)}{\sqrt{\zeta}}$
satisfies
$$|f''(\zeta)|=|-\ds\frac{\sin(\sqrt{\zeta}|x|)\sin(\sqrt{\zeta}|y|)(|x|^2+|y|^2)}
{4\zeta\sqrt{\zeta}}+\ds\frac{2\cos(\sqrt{\zeta}|x|)\cos(\sqrt{\zeta}|y|)|x||y|}
{4\zeta\sqrt{\zeta}}
$$
$$
-\ds\frac {\sin(\sqrt{\zeta}|x|)\cos(\sqrt{\zeta}|y|)|y|+
\cos(\sqrt{\zeta}|x|)\sin(\sqrt{\zeta}|y|)|x|}{2\zeta^2}
|\le\ds\frac{C(1+|x|^2)(1+|y|^2)}{\zeta\sqrt{\zeta}}.
$$
For  $y<0$ an identical calculation leads to the same bound.
Therefore the operator valued function ${\cal F}(\zeta)=\Gamma^+_{11}(\lam+0)-
\Gamma^+_{11}(\lam-0)$
satisfies the conditions (\ref{Z1}) and (\ref{Z2}) of Lemma \re{K} with
$\zeta=-\om-i\lam$, $p=1/2$ and ${\bf B}={\cal B}$.~~\bo
\\
Next we consider the second summand in the RHS of (\ref{Z}).
\begin{lemma}\la{decayP}
   In the situation of Theorem \ref{TD}
   \be
      \int\limits_{{\cal C}_+\cup{\cal C}_-}
      e^{\lam t}(P(\lam+0)-P(\lam-0))~d\lam={\cal O}(t^{-3/2}),
   \ee
   in the norm ${\cal B}$.
\end{lemma}
{\bf Proof }  We consider only the integral over
${\cal C}_+$ and one component of the matrix $P$, for example,
$P_{11}$:
$$
P_{11}(\lam,x,y)=\ds\frac{(i\al-2k_-)e^{ik_+(|x|+|y|)}
+(-i\al+2k_+)e^{ik_-(|x|+|y|)}+i\beta (e^{ik_-|y|+ik_+|x|}-
e^{ik_+|y|+ik_-|x|})}{2i\al(k_++k_-)-4k_+k_-+\al^2-\beta^2}.
$$
Denote $\zeta=-\om-i\lam$, then
$k_+=\sqrt{\zeta},\;k_-=\sqrt{-2\om-\zeta}.$ A
Taylor expansion in $\sqrt\zeta$ as $\zeta\to 0,\quad \rIm\zeta\ge 0$ implies
$$
P_{11}(\lam,x,y)=P_0+P_1(x,y)\zeta^{1/2}+P_2(x,y){\cal O}(\zeta),
$$
where $|P_j(x,y)|\le C_j(1+|x|^j)(1+|y|^j),\;j=1,2.$
Therefore, if $\lam\in{\cal C}_+$ then
$${\cal F}(\zeta)=P_{11}(\lam+0)-P_{11}(\lam-0)
={\cal O}(\zeta^{1/2}),\quad \zeta\to 0$$
in the norm of ${\cal B}$.
Similarly, differentiating two times the function $P_{11}(\lam,x,y)$
in $\lam$, we obtain that
$${\cal F}''(\zeta)=-P''_{11}(\lam+0)+P''_{11}(\lam-0)
={\cal O}(\zeta^{-3/2}),\quad i,j=1,2,\quad \zeta\to 0$$
in the norm of ${\cal B}$.
Moreover,
${\cal F}''(\zeta)\sim \zeta^{-3/2}$ as $\zeta\to\infty.$
Therefore, the function ${\cal F}(\zeta)$
satisfies the conditions (\ref{Z1}) and (\ref{Z2}) of Lemma \re{K} with $p=1/2$
and ${\bf B}={\cal B}$.~~\bo
\setcounter{equation}{0}
\section{Bounds for small $t$}
\label{decsubsecs}
As a starting point for  the method of majorants
in Section \ref{sassec} we will need also
some estimates on the  group  $e^{{\bf C}t}$ for small $t$.
First note that the function $e^{{\bf C}t}\chi_0$
belongs to $C_b(\R,H^1)$. This follows from a theorem analogous to
Theorem \ref{locex}
for solutions $e^{{\bf C}t}\chi_0$
of the linearized equation (\ref{lin4}),  with
initial condition $\chi_0\in H^1$.
Moreover, energy and charge conservation imply that
\be\la{t-small}
  \Vert e^{{\bf C}t}\chi_0\Vert_{H^1}\le c\Vert\chi_0\Vert_{H^1},\quad t\in \R.
\ee
For a further application
in section \ref{ptsec} we need a bound for the action of $e^{\bC t}$
on the Dirac distribution $\de=\de(x)$.
Thus let $\chi_{\delta}(x,t)$ be the solution to the
linearized equation (\re{lin3}) with $\chi_{\delta}(x,0)=\delta(x)$
and $e^{\bC t}\de$ its real vector version.
Note that, by Theorem \re{TD}, we have
$e^{\bC t}\de\in C_b(\ve,\infty;L^\infty_{-\beta})$,
for every $\ve>0$, and $\beta\ge 2$. The next lemma gives the small
$t$ behaviour:

\begin{lemma}\label{delta}
The following bound holds
\be\la{delta1}
  \Vert e^{{\bf C}t}\delta\Vert_{L^{\infty}}=\cO(t^{-1/2}),\quad t\to 0.
\ee
\end{lemma}
{\bf Proof }
By the Duhamel representation for the solution to (\re{lin3}), we obtain
\be\la{delta2}
\chi_{\delta}(x,t)=W_{\om}(t)\delta-\int\limits_0^tds
\Bigl(a\chi_{\delta}(0,s)+b\rRe(\chi_{\delta}(0,s)\Bigr)
W_{\om}(t-s)
\delta
\ee
where
$a$ and $b$ are defined by (\re{aa1}), and
$W_{\om}(t)$ is the dynamical group
of the modified Schr\"odinger equation
\be\la{mod-eqn}
i\dot\chi(x,t)=-\chi''(x,t)+\om\chi(x,t).
\ee
Note that
\be\la{delta3}
 W_{\om}(t)\delta=\frac 1{\sqrt{4\pi it}}~e^{i\frac{x^2}{4t}-i\om t}
\ee
Therefore (\re{delta2}) becomes
\be\la{delta4}
\chi_{\delta}(x,t)=\frac 1{\sqrt{4\pi it}}~e^{i\frac{x^2}{4t}-i\om t}
-\int\limits_0^t \frac 1{\sqrt{4\pi i(t-s)}}~e^{i\frac{x^2}{4(t-s)}-i\om (t-s)}
\Bigl(a\chi_{\delta}(0,s)+b\rRe(\chi_{\delta}(0,s)\Bigr)ds
\ee
Denote $\varsigma(x,t)= \sqrt{t}~\chi_{\delta}(x,t)$.
Then
\be\la{delta5}
\varsigma(x,t)=\frac 1{\sqrt{4\pi i}}~e^{i\frac{x^2}{4t}-i\om t}
-\sqrt{t}\int\limits_0^t \frac 1{\sqrt{4\pi i(t-s)s}}
~e^{i\frac{x^2}{4(t-s)}-i\om (t-s)}
\Bigl(a\varsigma(0,s)+b\rRe(\varsigma(0,s)\Bigr)ds.
\ee
Therefore,
\be\la{delta6}
\Vert\varsigma(t)\Vert_{L^{\infty}}\le\frac 1{2\sqrt{\pi}}
+\frac 12\sqrt{\pi t}(|a|+|b|)\int\limits_0^t \frac 1{\pi\sqrt{(t-s)s}}
~\Vert\varsigma(s)\Vert_{L^{\infty}}ds,~~~~~~~~t>0.
\ee
Since $\ds\int_0^t \frac {ds}{\pi\sqrt{(t-s)s}}=1$,
we obtain the bound
$$
\Vert\varsigma(t)\Vert_{L^{\infty}}\le\frac 1{2\sqrt{\pi}}\frac{1}
{1-\frac 12\sqrt{\pi t}(|a|+|b|)}
$$
if $t$ is sufficiently small.
\hfill\bo

\setcounter{equation}{0}
\section{Modulation equations }
\label{modsec}

In this section we present the modulation equations which allow a
construction of solutions $\psi(x,t)$ of the equation (\re{S})
close at each time $t$ to a soliton i.e. to one of the functions
$$
Ce^{i\theta-\sqrt{\om}|x|},\qquad\,\,C=C(\om)>0
$$
in the set $\cS$ described in section \ref{swsec} with time varying
(``modulating'') parameters $(\omega,\theta)=(\omega(t),\theta(t))$.
It will be assumed that $\psi(x,t)$ is a given weak solution of
(\re{S}) as provided by Theorem \ref{locex},
so that the map $t\to\psi(\,\cdot\,,t)$ is continuous into
$H^1(\R)$.
The modulation equations follow from the ansatz for the solution which
is explained next.
Recall that we defined
\be\la{exsol}
  \Phi_\om(x)\equiv \bigl(Ce^{-\sqrt{\om}|x|},0\bigr)=\bigl(\psi_\om,0\bigr)
\ee
so that $\psi(x,t)=e^{j\theta(t)}\Phi_{\om(t)}(x)$ is a solution of (\ref{SV}) if and
only if $\dot\theta=\om$ and $\dot\om=0$.
Here it is to be understood that $C=C(\om(t))$ is determined from $\omega(t)$
via (\ref{swac}).
We look for a solution to (\ref{SV}) in the form
\be\la{sol}
  \psi(x,t)=e^{j\theta(t)}\bigl(\Phi_{\om(t)}(x)+\chi(x,t)\bigr)=
  e^{j\theta(t)}\Psi(x,t),\quad \Psi(x,t)=\Phi_{\om(t)}(x)+\chi(x,t).
\ee
Since this is a solution of (\ref{SV}) as long as
$\chi\equiv 0$ and $\dot\theta=\om$ and $\dot \omega=0$
it is natural to look for solutions  in which $\chi$ is small and
$$\theta(t)=\int_0^t\om(s)ds+\gamma(t)$$ with $\gamma$ treated perturbatively.
Observe that so far this representation is underdetermined since for any
$\bigl(\om(t),\theta(t)\bigr)$ it just amounts to a
definition of $\chi$; it is made unique by restricting
$\chi(t)$ to lie in the image of the projection operator
onto the continuous spectrum ${\bf P}^c_t={\bf P}^c(\omega(t))$ or equivalently that
\be\la{req}
  \;{\bf P}^0_t \chi(t)=0,\;\;{\bf P}^0_t={\bf P}^0(\omega(t))=I-{\bf P}^c(\omega(t))
\ee
(The projection operators are as defined in (\ref{ILg}) and (\ref{defsp})).
An equivalent formulation of (\ref{req}) is to say that
$e^{j\theta}\chi$ is required to lie in the symplectic normal space
$N_{\omega(t),\theta(t)}\cS$.
This is equivalent to imposition of the following
orthogonality conditions (at each time $t$):
\be\la{orthog}
  \Om(\chi(t),T_0(\om(t))=0=\Om(\chi(t),T_1(\om(t)),
\ee
where $\Om$ is the symplectic form introduced previously. Writing $\chi(t)=(\chi_1(t),\chi_2(t))$
the orthogonality conditions reduce to
\be\la{orthog2}
  \int\, \chi_1(x,t) Ce^{-\sqrt{\om}|x|}\,dx=0,\quad\hbox{and}\quad
  \int\, \chi_2(x,t)\pa_\om (Ce^{-\sqrt{\om}|x|})\,dx=0.
\ee

Now we give a system of {\it modulation equations} for
$\omega(t),\;\gamma(t)$ which ensure the conditions (\ref{orthog2})
are preserved by the time evolution.
\begin{lemma}\label{BS}
  (i) Assume given a solution of (\ref{SV}) with regularity
  as described in theorem \ref{locex}, which can be
  written in the form (\ref{sol}) -(\ref{req}) with continuously differentiable
  $\omega(t),\;\theta(t)$. Then
  \be\la{SV2}
   \dot\chi={\bf C}\chi-\dot\om\pa_\om\Phi_\om+\dot\gamma j^{-1}(\Phi_\om+\chi)+{\bf Q}
  \ee
  where
  ${\bf Q}(\chi,\om)=-\delta(x) j^{-1}\bigl({\bF}(\Phi_\om+\chi)
  -{ \bF}(\Phi_\om)-{ \bF}'(\Phi_\om)\chi\bigr)$, and
  \beqn\la{omega}
    \dot\omega&=&\frac{\langle {\bf P}^0{\bf Q},\Psi\rangle}
    {\langle\pa_\om\Phi_{\om}-\pa_\om {\bf P}^0\chi,\Psi\rangle}\\
    \dot\gamma&=&\frac{\langle j{\bf P}^0(\pa_\om\Phi_{\om}-\pa_\om {\bf P}^0\chi),
    {\bf P}^0{\bf Q}\rangle}
    {\langle\pa_\om \Phi_\om-\pa_\om {\bf P}^0\chi,\Psi\rangle},\la{gamma}.
  \eeqn
  where ${\bf P}^0={\bf P}^0(\om(t))$ is the projection operator
  defined in (\ref{defsp}) and $\pa_\om{\bf P}^0=\pa_\om{\bf
  P}^0(\om)$ evaluated at $\om=\om(t)$.

  (ii) Conversely given $\psi$ a solution of (\ref{SV}) as in theorem \ref{locex} and
  continuously differentiable functions $\omega(t),\;\theta(t)$ which satisfy
  (\ref{omega})-(\ref{gamma}) then $\chi$ defined by (\ref{sol}) satisfies (\ref{SV2})
  and the condition (\ref{req}) holds at all times if it holds initially.
\end{lemma}
\Pr
This can be proved as in \ci[Prop.2.2]{BS}.\bo

It remains to show, for appropriate initial data close to a soliton, that there exist
solutions to (\ref{omega})-(\ref{gamma}), at least locally. To achieve this observe
that if the spectral condition (\ref{Co}) holds then by Lemma \ref{int-dif}
the denominator appearing on the right hand side of (\ref{omega}) and (\ref{gamma})
does not vanish for  small $\Vert\chi\Vert_{L^{1}_{\beta}}$. This is because
\be\la{int-dif1}
  \langle \pa_\om\psi_s,\psi_s\rangle=\frac{1}{2}\partial_\om\int|\psi_\om|^2dx\not=0
\ee
as discussed in section \ref{swsec}.
This has the consequence that the orthogonality conditions really can be satisfied for
small $\chi$ because they are equivalent to a locally well posed set of ordinary
differential equations for $t\to(\theta(t),\omega(t))$.
This implies the following corollary:
\begin{cor}\label{ME}
(i) In the situation of (i) in the previous lemma assume that
(\ref{int-dif1}) holds.
If $\Vert\chi\Vert_{L^p_\beta}$ is sufficiently small for some
$p,\beta$  the right hand sides of
(\ref{omega}) and (\ref{gamma}) are smooth in $\theta,\om$ and there exists
continuous ${\cal R}={\cal R}(\omega, \chi)$ such that
$$
  |\dot\gamma(t)|\le{\cal R}|\chi(0,t)|^2,\qquad
  |\dot\omega(t)|\le{\cal R}|\chi(0,t)|^2.
$$

(ii) Assume given $\psi$, a solution of (\ref{SV}) as in Theorem \ref{locex}.
If $\om_0$ satisfies (\ref{int-dif1}) and
$\chi(x,0)=e^{-j\theta_0}\psi(x,0)-\Phi_{\om_0}(x)$
is small in  some $L^p_\beta$ norm and satisfies (\ref{req}) there is a time interval
on which there exist $C^1$ functions $t\mapsto\bigl(\om(t),\gamma(t)\bigr)$ which
satisfy (\ref{omega})-(\ref{gamma}).
\end{cor}

\section{Time decay for the transversal dynamics}
\setcounter{equation}{0}
\label{sassec}
In this section we state our Theorem \ref{yest} on the time decay of the
transversal component $\chi(t)$ in the nonlinear setting, leaving the
proof to the next section. Theorem \ref{yest} will be
used to prove the main theorem in Section \ref{solas-sec}.
First we represent the initial data $\psi_0$ in a convenient form for
application of the modulation equations: the next Lemma will allow us
to assume that (\ref{req}) holds initially without loss of generality.
\begin{lemma}\la{in-cond}
In the situation of Theorem \ref{main} there exists a solitary wave
$\psi_{{\tilde\om}_0}={\tilde C}_0e^{-\sqrt{{\tilde\om}_0}|x|}$
satisfying the spectral condition (\ref{Co}) such that  in vector form
$$
\psi_0(x)=e^{j{\tilde\theta}_0}(\Phi_{{\tilde\om}_0}(x)+\chi_0(x)),
\quad  \Phi_{{\tilde\om}_0}=( \psi_{{\tilde\om}_0},0),
$$
and for $\chi_0(x)$ we have
 \be\la{innit}
   {\bf P}^0({\tilde\omega}_0)(\chi_0)=0,
\ee
and
  $$
    \Vert\chi_0\Vert_{L^1_{\beta}\cap H^1}= \tilde d=O(d)\quad\hbox{as}\; d\to 0.
  $$
\end{lemma}
{\bf Proof} By (\re{orthog}), the condition
(\ref{innit}) is equivalent to the pair of equations
  $$
  \Omega(e^{-j{\tilde\theta}_0}\psi_0-\Phi_{{\tilde\om}_0},T_0(\tilde\om_0))=0,\quad
  \Omega(e^{-j{\tilde\theta}_0}\psi_0-\Phi_{{\tilde\om}_0},T_1(\tilde\om_0))=0,
  $$
  where $T_0(\om)=j\Phi_{\om}$,
  $T_1(\om)=\partial_\om\Phi_{\om}$.
  For $\psi_0$ sufficiently close (in $L^1_{\beta}$) to $e^{j\theta_0}\Phi_{\om_0}$
  the existence of $\tilde\theta_0,\tilde\om_0$ follows by a standard application
  of the implicit function theorem  if we show that the Jacobian matrix
  \be\la{Jac}
  \left(\ba{ll}
  \partial_\om\Omega(e^{-j{\theta}}\psi_0-\Phi_{\om},j\Phi_{\om})&
  \partial_\om\Omega(e^{-j{\theta}}\psi_0-\Phi_{\om},\partial_\om\Phi_{\om})\\
  \partial_\theta\Omega(e^{-j{\theta}}\psi_0-\Phi_{\om},j\Phi_{\om})&
  \partial_\theta\Omega(e^{-j{\theta}}\psi_0-\Phi_{\om},\partial_\om\Phi_{\om})
  \ea\right),
  \ee
  with
$\psi_0=e^{j\theta_0}\Phi_{\om_0}$,
  is non-degenerate at $\om=\om_0$ and $\theta=\theta_0$.
  But this is equivalent to  the non-degeneracy of the matrix
  \be\la{Jac1}
  \left(\ba{cc}
  \Omega(\partial_\om\Phi_{\om_0},j\Phi_{\om_0}) &  0 \\
        0     &           \Omega(j\Phi_{\om_0},\partial_\om\Phi_{\om_0})
  \ea\right)
  \ee
  which holds by (\ref{int-dif1}).~~~~~~~~~~~~~~~~~~~~~~~~~~\bo

In Section \ref{solas-sec} we will show that our main Theorem \ref{main}
can be derived from the following time decay of the transversal component $\chi(t)$:
\begin{theorem}\la{yest}
  Let all the assumptions  of Theorem \ref{main} hold.
  For $d$ sufficiently small there exist $C^1$ functions $t\mapsto\bigl(\omega(t),\gamma(t)\bigr)$
  defined for $t\ge 0$ such that the solution $\psi(x,t)$  of  (\ref{SV})
  can be written as in (\ref{sol}-\ref{req})
  with (\ref{omega}-\ref{gamma})  satisfied, and there exists a number
  $\ov M>0$,
  depending only on the initial data, such that
  \be\la{ovY}
  M(T)=\sup\limits_{0\le t\le T}[(1+t)^{3/2}\Vert\chi(t)\Vert_{L^\infty_{-\beta}}
  +(1+t)^3\bigl(|\dot\gamma|+|\dot\omega|\bigr)|]\le \ov M,
  \ee
  uniformly in $T>0$, and $\ov M=O(d)$ as $d\to 0$.
\end{theorem}

\begin{remarks}

(0) This theorem will be deduced from Proposition \ref{ind-arg} in the next section.

  (i)  Theorem \ref{locex} implies that the norms in the definition
  of $M$ are continuous functions of time (and so $M$ is also).

  (ii) The result holds also for negative time with appropriate changes since
  $\psi(x,t)$ solves (\re{S}) if and only if $\overline\psi(x,-t)$ does.

  (iii) The result implies in particular that
  $t^3|\dot\theta-\om|+t^3|\dot\om|\le C$, hence $\om(t)$ and
  $\theta(t)-t\om_{+}$ should converge as $t\to\infty$ while
  $\psi(x,t)-e^{j\theta(t)}\Phi_{\om(t)}(x)$ have limit zero
  in $L^\infty_{-\beta}({\R})$.

  (iv) The notation $\chi(t)$ indicates the function $x\mapsto \chi(x,t)$ as usual.
\end{remarks}

\setcounter{equation}{0}
\section{Proof of transversal decay}
\label{prsec}
\subsection{Inductive argument (proof of Theorem \ref{yest})}
Let us write the initial data  in the form
  \be\la{0ansatz}
    \psi_0(x)=e^{j\theta_0}(\Phi_{\om_0}(x)+\chi_0(x)).
  \ee
  with $d=\Vert\chi_0\Vert_{L^1_\beta\cap H^1}$ sufficiently small.
By Lemma \ref{in-cond} we can assume that  ${\bf P}^0(\om_0)(\chi_0)=0$
without loss of generality. Then the local existence asserted in Corollary \ref{ME} implies
the existence of an interval $[0,t_1]$ on which are defined $C^1$ functions
$t\mapsto\bigl(\omega(t),\gamma(t)\bigr)$ satisfying (\ref{omega})-(\ref{gamma})
and such that $M(t_1)=\rho$ for some $t_1>0$ and $\rho>0$. By continuity we can make $\rho$
as small as we like by making $d$ and $t_1$ small. The following Proposition
  is proved in section \ref{ptsec} below.
\begin{pro}\la{ind-arg}
In the situation of Theorem \ref{yest}
let $M(t_1)\le \rho$ for some $t_1>0$ and $\rho>0$. Then
there exist numbers
$d_1$ and $\rho_1$, independent of $t_1$, such that
\be\la{half}
M(t_1)\le \rho/2
\ee
if $d=\Vert\chi_0\Vert_{L^1_\beta\cap H^1}<d_1$ and $\rho<\rho_1$.
\end{pro}
Assuming the truth of  Proposition \ref{ind-arg}
for now Theorem
\ref{yest}
will follow from the next argument:\\
Consider the set ${\cal T}$ of $t_1\ge 0$ such that
$\bigl(\omega(t),\gamma(t)\bigr)$ are defined on $[0,t_1]$ and $M(t_1)\le\rho$.
This set is relatively closed by continuity. On the other hand,
(\ref{half}) and Corollary \ref{ME} with sufficiently small $\rho$ and $d$
 imply that
this set is also relatively open, and hence $\sup {\cal
  T}=+\infty$, completing the proof of Theorem \ref{yest}.\bo
\subsection{Frozen linearized equation}
\label{autsec}

A crucial part of the proof of Proposition \ref{ind-arg} is the estimation
of the first term in $M$, for which purpose it is necessary to make
use of the dispersive properties obtained in sections \ref{subspace} and
\ref{rpsec}. Rather
than study directly (\ref{SV2}), whose linear part is non-autonomous,
it is convenient (following \cite{BP1,BP2}) to introduce a second ansatz, a small
modification of (\ref{sol}), which leads to an autonomous linearized equation.
This new ansatz for the solution is
\be\la{ansatz2}
  \psi(x,t)=e^{j\theta}(\Phi_\om(x)+e^{-j(\theta-\tilde\theta)}\eta),\quad\hbox{where}\;
  \tilde\theta(t)=\om_1t+\theta_0,\;\theta_0=\theta_0\;\hbox{and}\,
  \om_1=\om(t_1)
\ee
so that, $\eta=e^{j(\theta-\tilde\theta)}\chi$ and
$\chi=e^{-j(\theta-\tilde\theta)}\eta$.
Since
$$\dot\chi=e^{-j(\theta-\tilde\theta)}\Bigl(\dot\eta-j(\om+\dot\gamma-\om_1)\eta\Bigr)$$
equation (\ref{SV2}) implies
\be\la{eqeta}
  \dot\eta=j^{-1}(\om_1-\om)\eta+e^{j(\theta-\tilde\theta)}
  {\bf C}\Bigl(e^{-j(\theta-\tilde\theta)}\eta\Bigr)
  +e^{j(\theta-\tilde\theta)}\Bigl(j^{-1}\dot\gamma\Phi_\om-\dot\om\partial_\om\Phi_\om
  +{\bf Q}[e^{-j(\theta-\tilde\theta)}\eta]\Bigr).
\ee
The matrices ${\bf C}$ and $e^{j\phi}$, where $\phi=\theta-\tilde\theta$, do not commute
hence we need the following lemma:
\begin{lemma}\la{dif}
  \be\la{dif1}
  {\bf C}e^{j\phi}-e^{j\phi}{\bf C}=\delta(x)b\sin\phi~\sigma,\;{\rm where}\;\sigma=
  \left(\ba{cc}
  1    &    0\\
   0   &   -1
   \ea\right),\;b=2a'C^2.
  \ee
\end{lemma}
{\bf Proof }
  $$
  {\bf C}e^{j\phi}-e^{j\phi}{\bf C}=
  \left(\ba{cc}
        0            &   {\bf D}_2\\
   -{\bf D}_1        &     0
  \ea\right)
  \left(\ba{cc}
  \cos\phi    &   -\sin\phi\\
   \sin\phi   &   \cos\phi
  \ea\right)-
  \left(\ba{cc}
  \cos\phi    &   -\sin\phi\\
   \sin\phi   &   \cos\phi
  \ea\right)
  \left(\ba{cc}
        0            &   {\bf D}_2\\
   -{\bf D}_1        &     0
  \ea\right)
$$
$$
  =\left(\ba{cc}
  ({\bf D}_2-{\bf D}_1)\sin\phi    &   0\\
                 0                 &   ({\bf D}_1-{\bf D}_2)\sin\phi
  \ea\right)=
  \left(\ba{cc}
  \delta(x)b\sin\phi        &            0\\
                 0            &   -\delta(x)b\sin\phi
  \ea\right). ~~~~\bo
$$
Using Lemma \ref{dif} we rewrite equation (\ref{eqeta}) as
$$
  \dot\eta=j^{-1}(\om_1-\om)\eta+{\bf C}\eta+e^{j(\theta-\tilde\theta)}
  \Bigl(-\delta(x)b\sin(\theta-\tilde\theta)\sigma\eta+j^{-1}\dot\gamma\Phi_\om
  -\dot\om\partial_\om\Phi_\om+{\bf Q}[e^{-j(\theta-\tilde\theta)}\eta]\Bigr).
$$
To obtain a perturbed {\it autonomous} equation we rewrite the first two terms
on the RHS by freezing the coefficients at $t=t_1$. Note that
$$j^{-1}(\omega_1-\omega)+{\bf C}={\bf C_1}-j^{-1}\delta(x)(V-V_1),$$
where $V=a+bP_1$, $V_1=V(t_1)$, and ${\bf C_1}={\bf C}(t_1)$.
The equation for $\eta$ now reads
\be\la{eta-eq}
  \dot\eta={\bf C_1}\eta-j^{-1}\delta(x)(V-V_1)\eta+e^{j(\theta-\tilde\theta)}
  \Bigl(-\delta(x)b\sin(\theta-\tilde\theta)\sigma\eta+j^{-1}\dot\gamma\Phi
  -\dot\om\partial_\om\Phi_\om+{\bf Q}[e^{-j(\theta-\tilde\theta)}\eta]\Bigr)
\ee
The first term is now independent of $t$; the idea is that if
there is sufficiently rapid convergence of $\om(t)$ as $t\to\infty$
the other remaining  terms are small {\it uniformly with respect to} $t_1$.
Finally the equation (\ref{eta-eq}) can be written in the following
{\it frozen form}
\be\la{SV3}
  \dot\eta={\bf C_1}\eta+{\bf f_1}
\ee
where
\be\la{f1}
  {\bf f_1}= -j^{-1}\delta(x)(V-V_1)\eta+e^{j(\theta-\tilde\theta)}
  \Bigl(-\delta(x)b\sin(\theta-\tilde\theta)\sigma\eta+j^{-1}\dot\gamma\Phi
  -\dot\om\partial_\om\Phi_\om+{\bf Q}[e^{-j(\theta-\tilde\theta)}\eta]\Bigr)
\ee
\begin{remark}
  The advantage of (\ref{SV3}) over (\ref{SV2}) is that it can be treated as a perturbed
  autonomous linear equation, so that the estimates from section \ref{subspace} can be used
  directly. The  additional terms in ${\bf f_1}$
  can be estimated as small uniformly in $t_1$: see lemma \ref{alest} below.
  This is the reason for introduction of the second ansatz (\ref{ansatz2}).
\end{remark}
\begin{lemma}
\la{alest}
In the situation of Proposition \ref{ind-arg} there exists $c>0$,
independent of $t_1$, such that for $0\le t\le t_1$
$$
|a(t)-a_1|+|b(t)-b_1|+|\theta(t)-\tilde\theta(t)|\le c\rho.
$$
\end{lemma}
\Pr
By (\ref{ovY})
\be\la{cor-prop}
\sup\limits_{0\le t\le t_1}(1+t^3)(|\dot\gamma(t)|+|\dot\om(t)|)\le M(t_1)=\rho.
\ee
Therefore
$$
|a(t)-a(t_1)|=|\int\limits_t^{t_1}\dot a(\tau)d\tau|\le c\Bigl(\sup\limits_{0\le\tau\le t_1}
(1+\tau^2)|\dot\om(\tau)|\Bigr)\int\limits_t^{t_1}\fr{d\tau}{1+\tau^2}\le c\rho,
$$
since $|\dot a(\tau)|\le c|\dot\om(\tau)|$.
The difference $|b(t)-b(t_1)|$ can be estimated similarly.
Next
\beqn\la{theta-dif}
\theta(t)-\tilde\theta(t)&=&\int_0^t\om(\tau)d\tau+\gamma(t)-\om(t_1)t-\gamma(0)=
\int_0^t(\om(\tau)-\om(t_1))d\tau + \int_0^t\dot\gamma(\tau)d\tau\\
\nonumber
&=&-\int_0^t\int_{\tau}^{t_1}\dot\om(s)dsd\tau + \int_0^t\dot\gamma(\tau)d\tau.
\eeqn
By (\ref{cor-prop}) the first summand in RHS of (\ref{theta-dif}) can be estimated as
$$
\int_0^t\int_{t_1}^{\tau}|\dot\om(s)|ds\,d\tau\le
\int_0^t\int_{\tau}^{t_1}(1+s)^{2+\ve}|\dot\om(s)|\fr1{(1+s)^{2+\ve}}ds\,d\tau
$$
$$
\le c\sup\limits_{0\le s\le t_1}(1+s)^{2+\ve}|\dot\om(s)|
\int_0^t\int_{\tau}^{t_1}\fr1{(1+s)^{2+\ve}}ds\,d\tau\le c\rho
$$
since the last integral is bounded for $t\in[0,t_1]$.
Finally, for the second summand  on the RHS of (\ref{theta-dif}) inequality (\ref{cor-prop})
implies
$$
|\int_0^t\dot\gamma(\tau)d\tau|\le c\sup\limits_{0\le\tau\le t_1}
(1+\tau^2)|\dot\gamma(\tau)|\int\limits_t^{t_1}\fr{d\tau}{1+\tau^2}\le c\rho
$$
 \bo

\subsection{Projection onto discrete and continuous spectral spaces}
\label{decompsec}

From sections \ref{subspace} and \ref{rpsec} we have information concerning
$U(t)=e^{{\bf C_1}t}$, in particular decay on the subspace orthogonal
to the (two dimensional) generalized null space. It is
therefore necessary to introduce a further decomposition to take
advantage of this. Recall, by
comparing (\ref{sol}) and (\ref{ansatz2})  that
\be
\la{rtt}
\eta=e^{j(\theta-\tilde\theta)}\chi\quad\hbox{and}\;\;\;{\bf P}^0_t\chi(t)=0
\ee
Introduce the symplectic projections ${\bf P}^0_1={\bf P}^0_{t_1}$ and
${\bf P}^c_1={\bf P}^c_{t_1}$ onto the discrete and continuous
spectral subspaces defined by the operator ${\bf C_1}$ and write, at
each time $t\in[0,t_1]$:
\be\la{sdec2}
  \eta(t)=g(t)+h(t)
\ee
with $g(t)={\bf P}^0_1\eta(t)$ and  $h(t)={\bf P}^c_1\eta(t)$. The
following lemma shows that it is only necessary to estimate $h(t)$.
\begin{lemma}\label{g}
  In the situation of Proposition \ref{ind-arg}, assume
  $$
   \sup\limits_{0\le t\le t_1}\bigl(
   |\omega(t)-\omega_{1}|+|\theta(t)-\theta_{1}(t)|\bigr)=\Delta
  $$
  is sufficiently small. Then for $0\le t\le t_1$ there exists a linear operator
  $\Xi(t)$, bounded on $L^{\infty}_{-\beta}\cap H^1$,  and
  $c(\Delta,\omega_1)$ such that $\eta(t)=\Xi(t) h(t)$, and
  \be\la{chi}
c(\Delta,\omega_1)^{-1}|\Vert h\Vert _{L^{\infty}_{-\beta}\cap H^1}\le     \Vert \eta\Vert_{L^{\infty}_{-\beta}\cap H^1}\le c(\Delta,\omega_1)|\Vert h\Vert _{L^{\infty}_{-\beta}\cap H^1}.
  \ee
\end{lemma}
\Pr
  Explicitly we write
  \be\la{ndec}
     \eta(t)=h(t)+g(t),\qquad g(t)=b_0(t) T_0(\om_1)+b_1(t) T_1(\om_1)
  \ee
  where $b_0,b_1$ are chosen at each time $t$ to ensure that $\Omega(h(t),
  T_0(\om_1))=\Omega(h(t), T_1(\om_1))=0$.
  Using the fact that (since ${\bf P}^0_t\chi(t)=0$)
  $$
  \Omega\bigl(e^{-j(\theta-\tilde\theta)}\eta,T_0(\om(t)\bigr)=0=
  \Omega\bigl(e^{-j(\theta-\tilde\theta)}\eta,T_1(\om(t)\bigr)
  $$
  this means that $b_0,b_1$ are determined by
  \beqn\la{non}
    -\bmuone b_0(t)&=\Omega\bigl(\eta(t),
    T_1(\omega_1)\bigr)=\Omega\bigl(\eta(t),
    T_1(\omega_1)-e^{j(\theta-\tilde\theta)}T_1(\om(t))\bigr)\\
    \bmuone b_1(t)&=\Omega\bigl(\eta(t),
    T_0(\omega_1)\bigr)=\Omega\bigl(\eta(t),
    T_0(\omega_1)-e^{j(\theta-\tilde\theta)}T_0(\om(t))\bigr)\la{non'}.
  \eeqn
  From these it follows that there exists $c>0$ such that
  $\Vert g(t)\Vert_{L^{\infty}_{-\beta}\cap H^1}
  \le c\Delta\Vert\eta(t)\Vert_{L^{\infty}_{-\beta}\cap H^1}$
  and hence (\ref{chi}) follows
as claimed. \bo
\subsection{Proof of Proposition \ref{ind-arg}}
\label{ptsec}
To prove Proposition \ref{ind-arg}
we explain how to estimate both terms in $M$,
(\re{ovY}),
 to be $\le \rho/4$,
uniformly in $t_1$.\\
{\it  Estimation of the second term in $M$.}
As in Corollary \ref{ME} we have
$$
  |\dot\gamma(t)|+|\dot\om(t)|\le c_0|\chi(0,t)|^2\le c_0\fr{M(t)^2}{(1+|t|)^3},
  \quad t\le t_1,
$$
since $|\chi(0,t)|\le\Vert\chi(t)\Vert_{L^\infty_{-\beta}}$.
Finally let $\rho_1<1/(4c_0)$ to complete the estimate for the second term in $M$ as
$\le\rho/4$.\\
\noindent
{\it  Estimation of the first term in $M$.}
By Lemma \ref{g} it is enough  to estimate $h$.
Let us apply the projection ${\bf P}^c_1$ to both sides of  (\ref{SV3}).
Then the equation for $h$ reads
\be\la{h-eq}
  \dot h={\bf C_1}h+{\bf P^c_1f_1}
\ee
Now to estimate $h$ we use the Duhamel representation:
\be\la{Duh-rep}
h(t)=U(t) h(0)+\int_0^t U(t-s){\bf P}^c_1{\bf f_1}(s) ds,\quad t\le t_1.
\ee
with $U(t)=e^{{\bf C_1}t}$ the one parameter group just introduced.
Recall that
$
{\bf P}^0_1 h(t)=0$ for $t\in [0,t_1]$.
Therefore
\be\la{UU}
\Vert U(t)h(0)\Vert _{L^{\infty}_{-\beta}}\le c(1+t)^{-3/2}
\Vert h(0)\Vert_{L^1_{\beta}\cap H^1}\le c(1+t)^{-3/2}
\Vert \eta(0)\Vert_{L^1_{\beta}\cap H^1}.
\ee
by  Theorem \ref{TD} and inequalities (\ref{t-small}) and (\ref{chi}).
Let us estimate the integrand on the
right-hand side of (\ref{Duh-rep}).
\begin{lemma}\la{int-estimate}
The integrand in (\ref{Duh-rep}) satisfies the following bound:
\be\la{int-est}
\Vert U(t-s){\bf P}^c_1{\bf f_1(s)}\Vert _{L^{\infty}_{-\beta}}
\le c \frac {1}{(t-s)^{1/2}(1+t-s)}
\biggl(\Vert\eta(s)\Vert^2_{L^\infty_{-\beta}}
  +\rho\Vert\eta(s)\Vert_{L^\infty_{-\beta}}\biggr),~~~~~~0<s<t.
\ee
\end{lemma}
{\bf Proof}
We consider two different cases :   $t-s>\nu$, and  $0<t-s<\nu$, where $\nu=\nu(a,b)$ is defined in Lemma \re{delta}.\\
{\bf i) $t-s>\nu$ :}
We
use the representation (\re{f1})
for $\bf f_1$
and
apply   Theorem \ref{TD},
Corollary \ref{ME} and Lemma \ref{alest}  to obtain that
\beqn\la{U1}
\Vert U(t-s){\bf P}^c_1{\bf f_1}\Vert _{L^{\infty}_{-\beta}}
&\le& c(\nu)(1+t-s)^{-3/2}
\Vert {\bf P}^c_1({\bf f_1}(t))\Vert_{\M}\nonumber\\
&\le& c(\nu)(1+t-s)^{-3/2} \biggl(|\eta(0,t)|^2+\rho|\eta(0,t)|\biggr)\nonumber\\
&\le& c(\nu)(1+t-s)^{-3/2}
\biggl(\Vert\eta(t)\Vert^2_{L^\infty_{-\beta}}
  +\rho\Vert\eta(t)\Vert_{L^\infty_{-\beta}}\biggr),\quad t\le t_1.
\eeqn
{\bf ii) $0<t-s<\nu$ :}
We  denote ${\bf Q}=\delta(x){\bf\tilde Q}$, and represent ${\bf f_1}(x,s)$ as
\be\la{f1-dec}
{\bf f_1}(x,s)=p(s)\delta(x)+q(s)\Phi_{\om}+r(s)\partial_{\om}\Phi_{\om}.
\ee
where
$$
p(s)=-j^{-1}(V-V_1)\eta (0,s)+e^{j(\theta-\tilde\theta)}
  \Bigl(b\sin(\theta-\tilde\theta)\sigma\eta (0,s)+
  {\bf\tilde Q}[e^{-j(\theta-\tilde\theta)}\eta (0,s)]\Bigr),\;
$$
is an $\R^2$ valued function of time, and
$$
q(s)=-e^{-j(\theta-\tilde\theta)}j\dot\gamma,\quad
r(s)= -e^{-j(\theta-\tilde\theta)}\dot\om
$$
are ($2\times 2$) matrix valued functions of time. Writing
$\Vert\,\cdot\,\Vert$ for both the standard Euclidean and
operator norms on these, we have,
by Lemma \ref{alest},
$$
\Vert p(s)\Vert\le c\biggl(|\eta(0,s)|^2+\rho|\eta(0,s)|\biggr)
\le c \biggl(\Vert\eta(s)\Vert^2_{L^\infty_{-\beta}}
  +\rho\Vert\eta(s)\Vert_{L^\infty_{-\beta}}\biggr)
$$
and by Corollary \ref{ME}
$$
  \Vert q(s)\Vert, \Vert r(s)\Vert\le c|\eta(0,s)|^2
  \le c \Vert\eta(s)\Vert^2_{L^\infty_{-\beta}}.
$$
Applying projector ${\bf P}^c_1$ to  ${\bf f_1}$ we obtain
\be\la{Pf-dec}
{\bf P}^c_1{\bf f_1}(x,s)
=p(s)\delta(x)+q(s)\Phi_{\om} +r(s)\partial_{\om}\Phi_{\om}-{\bf P}^0_1{\bf f_1}(x,s).
\ee
By Lemma \ref{delta} for  sufficiently small  $\nu$ we obtain
\beqn\la{delta-est}
  &&\Vert U(t-s)p(s)\delta(x)\Vert_{L^{\infty}_{-\beta}}
  \le\Vert U(t-s)p(s)\delta(x)\Vert_{L^{\infty}}
  \le c(\nu)\Vert p(s)\Vert(t-s)^{-1/2}\nonumber\\
  &\le& c(\nu)(t-s)^{-1/2}\biggl(\Vert\eta(s)\Vert^2_{L^\infty_{-\beta}}
  +\rho\Vert\eta(s)\Vert_{L^\infty_{-\beta}}\biggr),\quad 0<t-s<\nu.
\eeqn
By inequality (\ref{t-small}) we have
\beqn\la{phi-est}
  &\Vert& U(t-s)\Bigl(q(s)\Phi_{\om}+r(s)\partial_{\om}\Phi_{\om}\Bigr)
  \Vert_{L^{\infty}_{-\beta}}
  \le c\Vert U(t-s)\Bigl(q(s)\Phi_{\om}+r(s)\partial_{\om}\Phi_{\om}\Bigr)\Vert_{H^1}\\
  \nonumber
  &\le& c \Bigl(\Vert q(s)\Vert\Vert\Phi_{\om}\Vert_{H^1}
  +(\Vert r(s)\Vert\Vert\partial_{\om}\Phi_{\om}\Vert_{H^1}\Bigr)
  \le c \Vert\eta(s)\Vert^2_{L^\infty_{-\beta}},\quad 0\le t-s<\nu.
\eeqn
The definition  (\ref{defsp}) of the projector ${\bf P}^0_1$
implies immediately that
$$
 \Vert {\bf P}^0_1{\bf f_1}\Vert_{H^1}
 \le c \Bigl(\Vert p(s)\Vert+\Vert q(s)\Vert +\Vert r(s)\Vert\Bigr)
$$
Then, similarly to (\ref{phi-est}), we obtain
\be\la{pr-est}
 \Vert U(t-s){\bf P}^0_1{\bf f_1}\Vert_{L^{\infty}_{-\beta}}
 \le c \bigl(\Vert\eta(s)\Vert^2_{L^\infty_{-\beta}}
  +\rho\Vert\eta(s)\Vert_{L^\infty_{-\beta}}\bigr),\quad 0\le t-s<\nu
\ee
Finally,  (\ref{Pf-dec})-(\ref{pr-est}) imply
\be\la{U-est}
\Vert U(t-s){\bf P}^c_1{\bf f_1}\Vert_{L^{\infty}_{-\beta}}
 \le c (t-s)^{-1/2}\bigl(\Vert\eta(s)\Vert^2_{L^\infty_{-\beta}}
  +\rho\Vert\eta(s)\Vert_{L^\infty_{-\beta}}\bigr),\quad 0<t-s<\nu .
\ee
From (\ref{U1}) and  (\ref{U-est}) inequality  (\ref{int-est}) follows.
\hfill\bo

Now (\ref{chi}), (\ref{Duh-rep}), (\ref{UU}) and (\ref{int-est}) imply
$$
\Vert\eta(t)\Vert_{L^\infty_{-\beta}}\le c(1+t)^{-3/2}\Vert\eta(0)\Vert_{L^1_{\beta}\cap H^1}
+c_1\int\limits_0^t \frac {ds}{(t-s)^{1/2}(1+t-s)}
\biggl(\Vert\eta(s)\Vert^2_{L^\infty_{-\beta}}
  +\rho\Vert\eta(s)\Vert_{L^\infty_{-\beta}}\biggr)
$$
Multiply by $(1+t)^{3/2}$ to deduce
\beqn
(1+t)^{3/2}\Vert\eta(t)\Vert_{L_{-\beta}^\infty}\le cd
&+&c_1\int\limits_{0}^t\frac{(1+t)^{3/2}(1+s)^{-3}}{(t-s)^{1/2}(1+t-s)}
(1+s)^{3}\Vert\eta(s)\Vert^2_{L^\infty_{-\beta}}ds \\
\nonumber
&+&c_1\rho\int\limits_{0}^t\frac{(1+t)^{3/2}(1+s)^{-3/2}}{(t-s)^{1/2}(1+t-s)}
(1+s)^{3/2}\Vert\eta(s)\Vert_{L^\infty_{-\beta}}ds
\eeqn
since $\Vert\eta(0)\Vert_{L^1_\beta\cap H^1}\le d$.
Introduce the majorant
$$
m(t):=\sup_{[0,t]}(1+s)^{3/2}\Vert\eta(s)\Vert_{L_{-\beta}^\infty},\quad t\le t_1
$$
and hence
\beqn\label{M}
m(t)\le cd+c_1 m^2(t)\int\limits_{0}^t
\frac{(1+t)^{3/2}(1+s)^{-3}}{(t-s)^{1/2}(1+t-s)}\,ds
+\rho c_1m(t)\int\limits_{0}^t
\frac{(1+t)^{3/2}(1+s)^{-3/2}}{(t-s)^{1/2}(1+t-s)}\,ds.
\eeqn
It easy to  see (by splitting up the integrals into
$s<t/2$ and $s\ge t/2$) that both these integrals are bounded independent of $t$.
Thus (\ref{M}) implies that there exist $c,c_2,c_3$, independent of
$t_1$, such that
$$m(t)\le cd+\rho c_2m(t)+c_3m^2(t),\quad t\le t_1.$$
Recall that $m(t_1)\le\rho\le\rho_1$ by assumption. Therefore
this inequality implies that $m(t)$ is bounded for $t\le t_1$, and moreover,
$$ m(t)\le c_4d,\quad t\le t_1$$
if $d$ and $\rho$ are sufficiently small. The constant $c_4$ does
not depend on $t_1$. We choose $d$ in (\ref{close}) small enough that $d<\rho/(4c_4)$.
Therefore,
$$\sup_{[0,t_1]}(1+t)^{3/2}\Vert\eta(t)\Vert_{L_{-\beta}^\infty} < \rho/4$$
if $d$ and $\rho$ are sufficiently small. This bounds the first
term as $<\rho/4$ by (\ref{rtt}) and hence
$M(t_1)< \rho/2$, completing the proof of Proposition \ref{ind-arg}.
\bo
\setcounter{equation}{0}
\section{Soliton asymptotics}
\label{solas-sec}

Here we prove our main Theorem \ref{main} using the bounds (\ref{ovY}).
For the solution $\psi(x,t)$ to (\ref{S}) let us define the accompanying soliton as
$s(x,t)=\psi_{\om(t)}(x)e^{i\theta(t)}$, where  $\dot\theta(t)=\om(t)+\dot\gamma(t)$.
Then for the difference $z(x,t)=\psi(x,t)-s(x,t)$ we obtain easily from
 equations (\ref{S}) and (\ref{NEP})
\be\la{z}
i\dot z(x,t)=-z''(x,t)+\dot\gamma s(x,t)-i\dot\om\partial_{\om}s(x,t)
-\delta(x)\Bigl(F(\psi(x,t))-F(s(x,t))\Bigr).
\ee
Then
\be\la{z1}
z(t)=W(t)z(0)+\int\limits_0^t W(t-\tau)\Bigl[\dot\gamma s(\tau)
-i\dot\om\partial_{\om}s(\tau)
-\delta(x)\Bigl(F(\psi(0,\tau))-F(s(0,\tau))\Bigr)\Bigr]d\tau,
\ee
where $z(t)=z(\cdot,t)$, $s(t)=s(\cdot,t)$, and $W(t)$ is the dynamical group
of the free Schr\"odinger equation.
Since $\gamma(t)-\gamma_{+}$,
$\om(t)-\om_{+}=\cO(t^{-2})$, and  therefore
$\theta(t)-\om_{+}t-\gamma_{+}=\cO(t^{-1})$ for $t\to\infty$,
to establish the asymptotic behaviour (\ref{sol-as}) it suffices to
prove that
\be\la{td}
z(t)=W(t)\Phi_{+}+r_{+}(t)
\ee
with some $\Phi_{+}\in C_b(\R)\cap L^2(\R)$ and
$\Vert r_{+}(t)\Vert_{C_b(\R)\cap L^2(\R)}=\cO(t^{-1/2})$.
Denote $g(t)=\dot\gamma s(t)-i\dot\om\partial_{\om}s(t)$,
$h(t)=F(\psi(0,t))-F(s(0,t))$ and rewrite equation (\ref{z1}) as
\be\la{z2}
z(t)=W(t)z(0)+W(t)\int\limits_0^tW(-\tau)g(\tau)d\tau
-W(t)\int\limits_0^tW(-\tau)\delta(x)h(\tau)d\tau.
\ee
Let us recall that $\dot\om(t),\,\dot\gamma(t)\sim t^{-3}$ as $t\to\infty$.
Hence,
for the second summand in RHS of (\ref{z2}) we have
\be\la{z3}
  W(t)\int\limits_0^t W(-\tau)g(\tau)d\tau
  =W(t)\int\limits_0^{\infty} W(-\tau)g(\tau)d\tau
  -W(t)\int\limits^{\infty}_t W(-\tau)g(\tau)d\tau =W(t)\phi_1+r_1(t),
\ee
where, from the unitarity in $H^1$ of the group $W(t)$ and the $t^{-3}$ decay of
$\dot\om$ and $\dot\gamma$, we infer that
 $\phi_1=\ds\int\limits_0^{\infty} W(-\tau)g(\tau)d\tau\in H^1$, and
 $\Vert r_1(t)\Vert_{H^1}=\cO(t^{-2}),\;t\to\infty$.\\
Consider now the  last summand in the RHS of (\ref{z2}).
Note that
$ W(t)\delta(x)=\ds\frac{e^{ix^2/(4t)}}{\sqrt{4\pi it}}$,
and
$|h(t)|\le c|\chi(0,t)|\le c {(1+t)^{-3/2}}$ by (\ref{ovY}).
Therefore
\be\la{z4}
W(t)\!\int\limits_0^t\!W(-\tau)\delta(x)h(\tau)d\tau =
W(t)\!\int\limits_0^{\infty}\frac {e^{-ix^2/(4\tau)}}{\sqrt{-4\pi i\tau}}~h(\tau)d\tau
-
\int\limits_t^{\infty}\frac {e^{ix^2/(4(t-\tau))}}{\sqrt{4\pi i(t-\tau)}}~h(\tau)d\tau
=W(t)\phi_2+r_2(t).
\ee
Evidently,
$\phi_2=\ds\int\limits_0^{\infty}\frac {e^{-ix^2/(4\tau)}}{\sqrt{-4\pi i\tau}}
~h(\tau)d\tau\in C_b$,
and $\Vert r_2(t)\Vert_{C_b}=\cO(t^{-1}),\;t\to\infty$.\\
Moreover,
$\phi_2\in L^2$, and  $\Vert r_2(t)\Vert_{L^2}=\cO(t^{-1/2}),\;t\to\infty$.
To see
that this is indeed true change variable to $\tau=1/u$ in the
definition to get:
\be
\la{ftz}
\phi_2(x)=\frac 1{\sqrt{-4\pi i}}
\int_0^\infty e^{-iux^2/4}~\eta(u)~du,\qquad \eta(u)=h(1/u)/u^{3/2}.
\ee
Now $h(t)$ is bounded and it follows from the decay of $h(t)$ that
$\eta(u)$ is bounded as $u\to 0$. Therefore $\eta(u)$
is square integrable and so by the Parseval-Plancherel theorem $\phi_2$
is square integrable as a function of $y=x^2$, and hence also as a function
of $x$ (since $dy=2xdx$ and $\phi_2$ is a bounded continuous function).
Next we have $r_2(t)=-W(t)R(t)$
with
$$
R(x,t)= \frac 1{\sqrt{-4\pi i}}
\int_0^{1/t} e^{-iux^2/4}~\eta(u)~du
=\frac 1{\sqrt{-4\pi i}}F_{u\to x^2/4} \zeta_t(u)\eta(u),
$$
where $\zeta_t(u)$ is the
characteristic function of the interval $(0,1/t)$.
The function $\eta(u)$ is bounded, hence $\Vert \zeta_t\eta\Vert_{L^2}=ct^{-1/2}$
and therefore $\Vert r_2(t)\Vert_{L^2}=\cO(t^{-1/2}),\;t\to\infty$.
To conclude,
using (\ref{z2}), (\ref{z3}), and (\ref{z4}) we obtain (\ref{td}) with
$\phi_{+}=z(0)+\phi_1+\phi_2$ and $r_{+}(t)= r_1(t)+r_2(t)$. The
$t\to -\infty$ case is handled in an identical way.
~~~\bo
\br
The expression (\ref{ftz}) for $\phi_2$ as a Fourier transform
implies immediately that $|\phi_2|$, and hence $|\Phi_+|$ also, tend to
0 as $|x|\to\infty$ by the Riemann-Lebesgue lemma. This same expression
could
be used with Zygmund's lemma to obtain more detailed decay properties of $\phi_2$ and hence
of $\Phi_+$. The decay rate would be determined essentially by the
regularity of the function $h(t)$ in addition to the decay rate of the
initial data.
\er

\setcounter{section}{0}
\setcounter{equation}{0}
\protect\renewcommand{\thesection}{\Alph{section}}
\protect\renewcommand{\theequation}{\thesection. \arabic{equation}}
\protect\renewcommand{\thesubsection}{\thesection. \arabic{subsection}}
\protect\renewcommand{\thetheorem}{\Alph{section}.\arabic{theorem}}

\section{The resolvent}
\subsection*{A.1 Calculation of the matrix kernel}
\label{res}
The derivation of the time decay of the solution to the
linearized equation (\re{lin3})
in section \ref{subspace} required an analysis of the smoothness and
singularities of the resolvent ${\bf R}(\lam)$ and its asymptotics
for $\lam\to\infty$. Here we will construct its matrix integral kernel
explicitly
\be\la{mic}
   {\bf R}(\lam,x,y)= \left( \ba{cc}R_{11}(\lam,x,y)&R_{12}(\lam,x,y)\\
   R_{21}(\lam,x,y)&R_{22}(\lam,x,y) \ea\right)
\ee
which is the solution to the equation
\be\la{mice}
   ({\bf C}-\lam){\bf R}(\lam,x,y)=\de(x-y)
   \left( \ba{cc}
   1&0\\
   0&1
   \ea\right).
   \ee
{\bf Calculation of first column}
For the first column
$R_I(\lam,x,y):= \left(\ba{c}R_{11}(\lam,x,y)\\
R_{21}(\lam,x,y) \ea\right)$  of the matrix ${\bf R}(\lam,x,y)$ we obtain
\be\la{mice1}
   ({\bf C}-\lam)R_I(\lam,x,y)=\de(x-y)
    \left( \ba{c}
   1\\
   0 \ea\right).
   \ee
If $x\ne 0$ and
$x\ne y$,  (\ref{mice1}) takes the form (cf.  (\re{D}), (\re{lin4}))
\be\la{hom}
   \left(
   \ba{lcl}-\lam  &&  {\bf D}_2\\
   -{\bf D}_1           &&  -\lam \ea \right)R_I(\lam,x,y) =\left(
   \ba{rcr}-\lam              &&-\ds\fr{d^2}{dx^2}+\om\\
   \ds\fr{d^2}{dx^2}-\om    &&-\lam \ea \right)R_I(\lam,x,y) =
   0,~~~~~x\ne 0,~~x\ne y.
\ee
The general solution is a linear combination of exponential solutions of type $e^{ikx}v$.
Substituting into (\re{hom}), we get
\be\la{homh}
   \left(
   \ba{rcr}-\lam   &&  k^2+\om\\
   -k^2-\om        &&    -\lam
   \ea
   \right)v=0.
\ee
For nonzero vectors $v$, the determinant of the matrix vanishes,
\be\la{det}
   \lam^2+(k^2+\om)^2=0.
\ee
Then $k_\pm^2+\om=\mp i\lam$.
Finally, we obtain four roots $\pm k_\pm(\lam)$ with
\be\la{ik}
   k_\pm(\lam)=\sqrt{-\om\mp i\lam},
\ee
where the square root is defined as an analytic continuation
from a neighborhood of the zero point $\lam=0$
taking the positive value of $\rIm\sqrt{-\om}$ at $\lam=0$.
We choose the cuts in the complex plane $\lam$ from the branching points to
infinity: the cut ${\cal C}_+:=[i\om,i\infty)$ for $k_+(\lam)$
and the cut ${\cal C}_-:=[-i\om,-i\infty)$ for $k_-(\lam)$. Then
\be\la{re}
   \rIm k_\pm(\lam)>0,~~~~~~~\lam\in\C\setminus {\cal C}_\pm.
\ee
It remains to derive the vector $v=(v_1,v_2)$ which is solution to (\re{homh}):
$$
   v_2=-\fr{k_\pm^2+\om}{\lam}v_1=\fr{ \pm i\lam}{\lam}v_1=\pm i v_1.
$$
Therefore, we have two corresponding vectors
$v_\pm=\left(
\ba{r}1\\
\pm i
\ea
\right)$
and we get four linearly independent exponential solutions
$$
   v_+e^{\pm ik_+ x}=\left(\ba{r}1\\i\ea\right)e^{\pm ik_+ x},~~~~~~~~~~~~
   v_-e^{\pm ik_- x}=\left(\ba{r} 1\\-i\ea\right)e^{\pm ik_- x}.
$$
Now we can solve the equation (\re{mice1}).
First we rewrite it using the representations (\re{lin4}) and (\re{D})
for the operator ${\bf C}$,
\be\la{homr}
   \!\left(\!
   \ba{rr}-\lam               &-\ds\fr{d^2}{dx^2}+\om\\
   \ds\fr{d^2}{dx^2}-\om   &-\lam
   \ea
   \!\right)\!\left(\ba{c}
   \!R_{11}(\lam,x,y)\\
   R_{21}(\lam,x,y) \ea\right) =\de(x-y) \left( \ba{c}
   1\\
   0
   \ea\!\right)+
   \de(x)
   \left(
   \!\ba{cc}
   0      &a\\
   -a-b &0
   \ea\!\right)\!\left(\ba{c}
   R_{11}(\lam,0,y)\\
   R_{21}(\lam,0,y)
   \ea\right)
\ee
Let us consider $y>0$ for the concreteness. Then
the RHS vanishes in the open intervals $(-\infty,0)$,$(0,y)$ and $(y,\infty)$.
Hence, for the parameter $\lam$ outside the cuts $C_\pm$,
the solution admits the representation
\be\la{ABC}
   R_{I}(\lam,x,y)
   =\left\{ \ba{ll}
   A_{+}e^{-ik_{+}x}v_{+} +A_{-}e^{-ik_{-} x}v_{-},  \!&\!x<0,\\\\
   B^{-}_{+}e^{-ik_{+} x}v_{+}+B^{-}_{-}e^{-ik_{-} x}v_{-}+B^{+}_{+}e^{ik_{+} x}v_{+}
   +B^{+}_{-}e^{ik_{-} x}v_{-},  \!&\!0<x<y,\\\\
   C_{+}e^{ik_{+} x}v_{+}+C_{-}e^{ik_- x}v_{-},  \!&\!x>y
   \ea
   \right.
\ee
since by (\re{re}), the exponent $e^{-ik_\pm x}$ decays for $x\to -\infty$,
and similarly, $e^{ik_\pm x}$ decays for $x\to\infty$.
Next we need eight equations to calculate the eight constants $A_+,\dots,C_-$.
We have two continuity equations and two jump conditions for the derivatives
at the points $x=0$ and $x=y$. These four vector equations
give just eight scalar equations for the calculation.
\\
{\bf Continuity at $x=y$:} $R_I(y-0,y)=R_I(y+0,y)$, i.e.
$$
  B^{-}_{-}v_{+}/e_{+}+B^{-}_{-}v_{-}/e_{-}+B^{+}_{+}v_{+}e_{+}+B^{+}_{-}v_{-}e_{-}
  =C_{+}v_{+}e_{+}+C_{-}v_{-}e_{-},
$$
where  $e_\pm:= e^{ik_\pm y}$.
It is equivalent to
\be\la{xy}
   \left\{
   \ba{l}
   B^{-}_{+}/e_{+}+B^{+}_{+}e_{+}=C_{+}e_{+},\\\\
   B^{-}_{-}/e_{-}+B^{+}_{-}e_{-}=C_{-}e_{-}.
   \ea
   \right.
\ee
{\bf Continuity at $x=0$:} $R_I(-0,y)=R_I(+0,y)$, i.e.
$$
   A_{+}v_{+}+A_{-}v_{-}=B^{-}_{+}v_{+}+B^{-}_{-}v_{-}+B^{+}_{+}v_{+}+B^{+}_{-}v_{-}
$$
that is equivalent to
\be\la{x0}
   \left\{
   \ba{l}
   A_{+}=B^{-}_{+}+B^{+}_{+},\\\\
   A_{-}=B^{-}_{-}+B^{+}_{-}.
   \ea
   \right.
\ee
{\bf Jump at $x=y$:}
 $R'_I(y+0,y)=R'_I(y-0,y)+
\left(
\ba{r}
0\\
-1
\ea\right)
$,
where prime denotes the derivative in $x$. Substituting (\re{ABC}), we get
\beqn\la{j0y}
   &&ik_{+}C_{+}v_{+}e_{+} +ik_{-}C_{-}v_{-}e_{-} =\nonumber\\\nonumber\\
   &&-ik_{+}B^{-}_{+}v_{+}/e_{+} -ik_{-}B^{-}_{-}v_{-}/e_{-}
   +ik_{+}B^{+}_{+}v_{+}e_{+}+ik_{-}B^{+}_{-}v_{-}e_{-}+
   \left(
   \ba{r}
   0\\
   -1
   \ea\right)
\eeqn
Noting that
\be\la{vpm}
   \left(
   \ba{r}
   0\\
   -1
   \ea\right)=\ds\fr{v_+-v_-}2i,
\ee
we get
\be\la{weg}
\left\{
\ba{c}
ik_{+}C_{+}e_{+}=-ik_{+}B^{-}_{+}/e_{+}+ik_{+}B^{+}_{+}e_{+} +\ds\fr i2,\\
\\
ik_{-}C_{-}e_{-}=-ik_{-}B^{-}_{-}/e_{-}+ik_{-}B^{+}_{-}e_{-}-\ds\fr i2.
\ea\right.
\ee
After substituting of $C_{\pm}$ from (\re{xy}), the constants
 $B^{+}_{\pm}$ cancel and we get
\be\la{BB}
B^{-}_{+}=\ds\fr {e_+}{4k_+},~~~~~~~~~~~~~~~~B^{-}_{-}=-\ds\fr{e_-}{4k_-}.
\ee
{\bf Jump at $x=0$:}
$R'_I(+0,y)=R'_I(-0,y)-
 \left(
 \ba{cc}
 a+b&0\\
 0&a
\ea\right)R_I(-0,y)$.
Substituting (\re{ABC}), we get
\be\la{j0}
-ik_{+}B^{-}_{+}v_{+}-ik_{-}B^{-}_{-}v_{-}+ik_{+}B^{+}_{+}v_{+}
+ik_{-}B^{+}_{-}v_{-}
=-ik_{+}A_{+}v_{+}-ik_{-}A_{-}v_{-}-M(A_{+}v_{+}+A_{-}v_{-}),
\ee
where $M$ is the matrix
$\left(\ba{cc}a+b&0\\0&a\ea\right)$.
Note that
\be
\left\{
\ba{c}
Mv_+=\al v_++\beta v_-\\\\
Mv_-=\al v_-+\beta v_+
\ea\right.,~~~~~~{\rm where}~~~~
\al=a+\fr{b}2,\quad\beta=\fr{b}2.
\la{defal}
\ee
Then (\re{j0}) becomes
$$
\left\{
\ba{l}
-ik_{+}B^{-}_{+}+ik_{+}B^{+}_{+}=-ik_{+}A_{+}-A_{+}\al-A_{-}\beta,
\\
\\
-ik_{-}B^{-}_{-}+ik_{-}B^{+}_{-}=-ik_{-}A_{-}-A_{+}\beta-A_{-}\al.
\ea\right.
$$
Substituting here (\re{x0}), we get after cancellations,
$$
\left\{
\ba{l}
(2ik_{+}+\al)B^{+}_{+}+\beta B^{+}_{-}=-\al B^{-}_{+}-\beta B^{-}_{-}
\\
\\
\beta B^{+}_{+}+(2ik_{-}+\al)B^{+}_{-}=-\beta B^{-}_{+}-\al B^{-}_{-}
\ea\right.
$$
Hence, the solution is given by
\be\la{BBB}
\left(
\ba{r}
B^{+}_{+}\\
B^{+}_{-}
\ea\right)
=-\fr 1{D}
\left(
\ba{cc}
2ik_{-}+\al&  -\beta\\
-\beta     &  2ik_{+}+\al
\ea\right)
\left(
\ba{cc}
\al&\beta\\
\beta&\al
\ea\right)
\left(
\ba{r}
B^{-}_{+}\\
B^{-}_{-}
\ea\right),
\ee
where $D$ is the determinant
\be\la{dete}
D:=(2ik_++\al)(2ik_-+\al)-\beta^2,
\ee
and $B^{-}_{+},B^{-}_{-}$ are given by (\re{BB}).
The formulas (\re{BB}) and (\re{BBB}) imply
\be\la{BBBB}
B^{+}_{+}=\frac 1{2D}\left(-\ds\frac{2ik_{-}\al+\al^2-\beta^2}{2k_+}e_{+}
+i\beta e_{-}\right),\quad
B^{+}_{-}=\frac 1{2D}\left(-i\beta e_{+}+\ds\frac{2ik_+\al+\al^2-\beta^2}{2k_-}e_{-}
 \right).
\ee
Using the identities
$$ 2ik_-\al+\al^2-\beta^2=D-2ik_+\al+4k_+k_-,\quad
   2ik_+\al+\al^2-\beta^2=D-2ik_-\al+4k_+k_-,$$
let us rewrite (\ref{BBBB}) as
\be\la{fin}
B^{+}_{+}=-\ds\frac{e_+}{4k_+}+\frac 1{2D}\Bigl((i\al-2k_{-})e_{+}+i\beta e_{-}\Bigr),
\quad
B^{+}_{-}=\ds\frac{e_-}{4k_-}-\frac 1{2D}\Bigl(i\beta e_{+}+(i\al-2k_{+})e_{-}\Bigr).
\ee
Finally, the formulas  (\re{ABC})--(\re{x0}),
(\re{BB}) and (\re{fin}) give the first column $R_I(\lam,x,y)$ of
the resolvent for $y>0$:
\be\la{RI}
R_{I}(\lam,x,y)=\Gamma_{I}(\lam,x,y)+P_{I}(\lam,x,y),
\ee
where
\be\la{GI}
\Gamma_{I}(\lam,x,y)=\frac 1{4k_+}(e^{ik_{+}|x-y|}-e^{ik_{+}(|x|+|y|)})v_{+}
-\frac 1{4k_-}(e^{ik_{-}|x-y|}-e^{ik_{-}(|x|+|y|)})v_{-},
\ee
and
\beqn\la{PI}
P_{I}(\lam,x,y)&=&\frac 1{2D}\Bigl[\Bigl((i\al-2k_{-})e^{ik_{+}(|x|+|y|)}
+i\beta e^{i(k_{+}|x|+k_{-}|y|)}\Bigr)v_{+}\\
\nonumber
&-&\Bigl(i\beta e^{i(k_{-}|x|+k_{+}|y|)}+(i\al-2k_{+})e^{ik_{-}(|x|+|y|)}
\Bigr)v_{-}\Bigr]
\eeqn
\\
{\bf Calculation of second column}
The second column is given by similar  formulas with the vector
$\left(
\ba{r}
1\\
0
\ea\right)
$
instead of
$\left(
\ba{r}
0\\
1
\ea\right)
$
in (\re{homr}). Then
$\left(
\ba{r}
0\\
-1
\ea\right)
$
in (\re{j0y}) is changed by
$\left(
\ba{r}
1\\
0
\ea\right)
$.
Respectively, (\re{vpm}) is changed by
$$
\left(
\ba{r}
1\\
0 \ea\right)=\ds\fr{v_-+v_+}{2}.
$$
Hence, we have now change $i/2$ by $1/2$ in the first equation of (\re{weg}) and $-i/2$
by $1/2$ in the second one. Respectively, (\re{BB}) for the second column reads
$$
B^{-}_{+}=-\ds\fr {ie_+}{4k_+},~~~~~~~~~~~~~~~~
B^{-}_{-}=-\ds\fr {ie_-}{4k_-}.
$$
Then the second column
$R_{II}(\lam,x,y)$ of the resolvent reads:
\be\la{RII}
R_{II}(\lam,x,y)=\Gamma_{II}(\lam,x,y)+P_{II}(\lam,x,y),
\ee
where
\be\la{GII}
\Gamma_{II}(\lam,x,y)=-\frac i{4k_+}(e^{ik_{+}|x-y|}-e^{ik_{+}(|x|+|y|)})v_{+}
-\frac i{4k_-}(e^{ik_{-}|x-y|}-e^{ik_{-}(|x|+|y|)})v_{-},
\ee
and
\beqn\la{PII}
P_{II}(\lam,x,y)&=&\frac i{2D}\Bigl[\Bigl(-(i\al-2k_{-})e^{ik_{+}(|x|+|y|)}
+i\beta e^{i(k_{+}|x|+k_{-}|y|)}\Bigr)v_{+}\\
\nonumber
&+&\Bigl(i\beta e^{i(k_{-}|x|+k_{+}|y|)}-(i\al-2k_{+})e^{ik_{-}(|x|+|y|)}
\Bigr)v_{-}\Bigr]
\eeqn
Note, that if $y<0$ we get the same formulas.
\subsection*{A.2 The poles of the resolvent}
\label{polres}
The poles of the resolvent correspond to the roots of the determinant (\re{dete}),
\be\la{deter}
  D(\lam):=\al^2+2i\al(k_++k_-)-4k_+k_--\beta^2=0.
\ee
with $k_\pm$ as in (\ref{ik})-(\ref{re}). Thus $D(\lam)$ is an analytic
function on $\C\setminus {\cal C}_-\cup{\cal C}_+$.
Since there are two possible
values for the square roots in $k_\pm$ there is a corresponding
four-sheeted function $\tilde D(\lam)$ analytic on a four sheeted
cover
of $\C$ which is branched over ${\cal C}_-$ and ${\cal C}_+$. We call
the sheet defined by (\ref{re}) the {\it physical sheet}.

We will reduce the equation (\re{deter}) to the solution of two
successive quadratic equations. These can be solved explicitly but
the process involves squaring and thus actually produces zeros
of the function $\tilde D(\lam)$ rather than of $D(\lam)$.
Therefore we will then have to check whether or not the roots do
actually lie on the physical sheet.
\\
{\it Step i)}\\
Denote by $\si=k_++k_-$. Then
\be\la{si2}  \si^2=2k_+k_--2\om
\ee
by (\re{ik}), hence (\re{deter}) gives the {\it first quadratic equation}:
$$
\al^2+2i\al \si-2(\si^2+2\om)-\beta^2=0.
$$
Rewrite it as
\be\la{detet}
  \si^2-i\al \si=\ds\fr {\al^2-\beta^2}2-2\om=:\de
\ee
Finally,
\be\la{detof}
\si=\ds\fr{i\al }2
\pm\sqrt{\de-\ds\fr{\al^2}4},
\ee
where the root is choosen arbitrarily.

Further let us express the roots in $\om$. Since $a=2\sqrt{\om}$, $\al=a+b/2$,
$\beta=b/2$ then substituting $\de$ from (\ref{detet}), we obtain
$$
\de-\ds\fr{\al^2}4=\ds\fr{\al^2}4-\ds\fr{\beta^2}2-2\om=
\ds\fr{(a+b/2)^2}{4}-\ds\fr{b^2}{8}-\ds\fr{a^2}2=-\ds\fr{a^2}4-\ds\fr{b^2}{16}
+\ds\fr{ab}4=-\ds\fr1{16}(2a-b)^2<0.
$$
Now (\re{detof}) reads
\be\la{detor}
 \si=\ds\fr{i\al }2\pm\ds\fr i4(2a-b)=\fr i4\Big[(2a+b)\pm(2a-b)\Big]=i\ga_j,\;j=1,2,
\ee
where  $\ga_j\in\R$, and
\be\la{gaga}
\ga_1=a=a(C^2),~~~~~~~~~\ga_2=b/2=a'(C^2)C^2.
\ee
\\
{\it Step ii)}\\
It remains to calculate the correponding spectral parameter $\lam$.
First, the quadratic equation (\re{si2}) implies by (\re{detor}) that
\be\la{si2i}
  4(k_+k_-)^2=(2\om+\si^2)^2=(2\om-\ga_j^2)^2,\;j=1,2.
\ee
On the other hand,
\be\la{kk}
  k_+k_-=\sqrt{-\om+i\lam}\sqrt{-\om-i\lam},
\ee
hence (\re{si2i}) gives the {\it second quadratic equation}
$$
4(\om^2+\lam^2)=(2\om-\ga_j^2)^2.
$$
Therefore,
$$
\lam^2=\ds\fr{(2\om-\ga_j^2)^2-4\om^2}4=
-\ds\fr{\ga_j^2(4\om-\ga_j^2)}4.
$$
Finally, we obtain four roots
\be\la{lam}
  \lam_j=i\ds\fr{\ga_j}2\sqrt{4\om-\ga_j^2},
\ee
where $j\in\{1,2\}$ and the square root can takes two opposite values.
\bc
\la{rootc}
The four-sheeted function $\tilde D(\lam)$ has the following roots (zeros):

i) $j=1$ gives $\lam_1=0$ since $4\om=a^2=\ga_1$.

ii) If $|\ga_2|<2\sqrt{\om}$, then both $j=2$ roots
$\pm i|\lam_2|$ are pure imaginary.

iii) If $|\ga_2|>2\sqrt{\om}$,
then  both $j=2$ roots $\pm|\lam_2|$ are real:
one positive and one negative.
\ec
\begin{remark} Note that a priori we can meet the wrong sign of $\rIm k_{\pm}$
squaring (\ref{kk}) which is why the above calculation yields roots of
$\tilde D(\lam)$ rather than the physical branch $D(\lam)$.
Since the formulas (\ref{RI})-(\ref{PII}) involve only
$D(\lam)$ it is important to know which of
these are actually roots of $D(\lam)$ and also to know the
multiplicities. This is done in the next two
sections.
\end{remark}
\subsection*{A.3 Discrete spectrum $\lam=0$}
\label{discrete1}

In order to check that the roots of $\tilde D(\lam)$
given in Corollary \ref{rootc}
are actually roots of $D(\lam)$ it suffices
to check explicitly that $D(\lam)$ vanishes
(with the assumption that we are on the physical branch defined by
$\rIm k_{\pm} > 0$ for
$\lam\in\C\setminus {\cal C_{\pm}}$.

For $j=1$ we have $\ga=\ga_1=a=2\sqrt\om$ and then $\lam_1=0$.
For $j=2$ we have  $\ga=\ga_2=a'C^2$.
If $|\ga_2|=2\sqrt{\om}$ ( equivalently  $|a'|=a/C^2$)
or $\ga_2=0$ ( equivalently $a'=0$),
we have $\lam_2=0$.

Let us check that $\lam=0$ is a root of $D(\lam)$:
$$
  D(0)=\alpha^2-\beta^2+2i\alpha 2i\sqrt{\om}+4\om=(a+b/2)^2-b^2/4-2(a+b/2)a+a^2=0
$$
since  $k_{\pm}=i\sqrt{\om}$.
Now let us compute $D'(\lam)$:
$$
D'(\lam)=i\alpha(\fr{i}{\sqrt{-\om+i\lam}}+\fr{-i}{\sqrt{-\om-i\lam}})
-(\fr{2i}{\sqrt{-\om+i\lam}}\cdot\sqrt{-\om-i\lam}+
\fr{-2i}{\sqrt{-\om-i\lam}}\cdot\sqrt{-\om+i\lam}).
$$
Hence $D'(0)=0$ and $\lam=0$ is the root of $D(\lam)$ of multiplicity at least 2. Further
calculation shows that the Taylor series for $D$ near zero takes the form:
\be
\label{tayld}
D(\lam)=\bigl(\frac{1}{\om}-\frac{b}{4\om^{3/2}}\bigr)\lam^2+O(\lam^4).
\ee
Therefore $\lam=0$ is the root of $D(\lam)$ of multiplicity 4 if and only if
$b=4\sqrt\om$, i.e. $a'=a/C^2$, and we have proved the following lemma:
\begin{lemma}\la{lam0}
If $a'=a/C^2$ then $\lam=0$ is a root of the determinant $D(\lam)$ with
multiplicity 4, otherwise $\lam=0$ is a root of the determinant
$D(\lam)$ with multiplicity 2.
\end{lemma}
\subsection*{A.4 Nonzero discrete spectrum }
\label{discrete2}

Now let us check whether the roots $\lam=\lam_2\not=0$ corresponding
$\ga=\ga_2\not\in\{0,\pm 2\sqrt{\om}\}$ lie on the physical branch.
We analyze two different cases: $0<|\ga_2|<2\sqrt\om$ and $|\ga_2|>2\sqrt\om$.
\\
{\bf I.The  case $\bf{0<|\ga_2|<2\sqrt\om}$} (equivalently $0<|a'|< a/C^2$).
\\
Since $4\om-\ga_2^2>0$, the corresponding roots $\lam_2$ are pure imaginary by (\re{lam}).
Moreover, $|\lam_2|\le\om$.
Indeed, (\re{lam}) implies
$$
\om^2-|\lam_2|^2 =\om^2+\ga_2^4/4-\ga_2^2\om=(\om-\ga_2^2/2)^2\ge0.
$$
Hence  $-\om\mp i\lam_2\le 0$  and
$k_{\pm}$ are pure imaginary with nonnegative imaginary part, that is
\be\label{con}
   k_+k_-\le 0\;\rm {and}\;\rIm(k_++k_-)>0.
\ee
The equations (\ref{si2i}) and  (\ref{si2}) imply
\be\label{con1}
|k_+k_-|=\fr14|a^2-2(a')^2C^4|,\,\,\,(k_++k_-)^2=-2\om+2k_+k_-=-\fr{a^2}{2}+2k_+k_-.
\ee
In order to obtain $k_+k_-$ and $k_++k_-$ from the last two equations
we have to divide the set $0<|a'|< a/C^2$ onto three subsets:
$$
(-a/C^2, a/C^2)\setminus\{0\}=(-a/C^2,-a/\sqrt{2}C^2]\cup \Big((-a/\sqrt{2}C^2,a/\sqrt{2}C^2)\setminus
\{0\}\Big)
\cup [\frac a{\sqrt{2}C^2},\frac a{C^2}) .
$$
1) First consider the case $a'\in [\ds\frac a{\sqrt{2}C^2},\ds\frac a{C^2})$.
Then (\ref{con}) and (\ref{con1}) imply
$$
k_+k_-=\fr 14(a^2-2(a')^2C^4).
$$
$$
(k_++k_-)^2=-\fr {a^2}{2}+\fr {a^2}{2}-(a')^2C^4=-(a')^2C^4,
$$
$$
k_++k_-=ia'C^2,
$$
and using (\ref{deter}), we obtain
$$
D(\lam_2)=(a+a'C^2)^2-(a'C^2)^2+2i(a+a'C^2)(k_++k_-)-4k_+k_-
$$
$$
=a^2+2aa'C^2-2(a+a'C^2)a'C^2-a^2+2(a')^2C^4=0.
$$
Note that each $\ga_2$ defines two values $\lam_2$
up to factor $\pm 1$.
If we replace $\lam_2$ by $-\lam_2$,
 $k_+$ and $k_-$ change places and our calculation remains valid.
Therefore, both values of $\lam_2$ are roots of $D(\lam)$.

2) Further consider
$a'\in (-\ds\frac a{C^2},-\ds\frac a{\sqrt{2}C^2}]$. In this case
$$
k_+k_-=\fr14(a^2-2(a')^2C^4),\quad k_++k_-=-ia'C^2.
$$
Then we have
$$
D(\lam_2)=a^2+2aa'C^2+2(a+a'C^2)a'C^2-a^2+2(a')^2C^4
=4a'C^2(a+a'C^2)\not =0
$$
since $a'\not = 0$ and $a'\not=-a/C^2$.
Therefore in this case  both values of $\lam_2$ are not the roots of $D(\lam)$.
\\
3) Finally consider
$0<|a'|<\ds\frac a{\sqrt{2}C^2}$. Then
(\ref{con})-(\ref{con1} imply that
$$
k_+k_-=-\fr14(a^2-2(a')^2C^4)<0,
$$
$$
(k_++k_-)^2=-a^2+(a')^2C^4<0,
$$
$$
k_++k_-=i\sqrt{a^2-(a')^2C^4}.
$$
Then we have
\be\label{D3}
D(\lam_2)=a(a+2a'C^2)-2(a+a'C^2)\sqrt{a^2-(a')^2C^4}+a^2-2(a')^2C^4.
\ee
To solve the equation $D(\lam_2)=0$ with respect to $a'$,
divide the RHS of (\ref{D3}) by
$C^4\not =0$ and denote $p=a/C^2>0$. Then we get the equation
\be\label{D31}
p^2+pa'-(a')^2 =(p+a')\sqrt{p^2-(a')^2},\;0<|a'|< p/\sqrt 2.
\ee
Squaring  both side of (\ref{D31}), we get
$$
2(a')^4-p^2(a')^2 =0
$$
The equation has no solutions for $0<|a'|< p/\sqrt 2$
and hence $D(\lam_2)$ does not vanish.
\begin{cor}
i)
 $D(\lam_2)=0$
if $\ds a'\in [\frac a{\sqrt{2}C^2},\frac a{C^2})$.
\\
ii)
$D(\lam_2)\not =0$ if
$\ds a'\in (-\frac a{C^2},\frac a{\sqrt{2}C^2})\setminus\{0\}$.
\end{cor}
{\bf II. The case $\bf{|\ga_2|>2\sqrt{\om}}$ } (equivalently $|a'|>a/C^2$).
\\
Since $4\om-\ga_2^2<0$,
the corresponding roots (\re{lam}) are real:
$\lam_2^-<0<\lam_2^+$, $\lam_2^-=-\lam_2^+$.
It is easy to prove that $k_{\pm}$ take the form:
$$k_{\pm}=\pm\mu+i\nu,\quad\nu>0.$$
Therefore
\be\la{kpm}
k_{+}k_{-}=-\mu^2-\nu^2<0, \quad k_{+}+k_{-}=2i\nu
\ee
\\
1) First consider the case $a'>a/C^2$.Then  by (\ref{con1}) and (\ref{kpm})
$$
k_+k_-=\frac 14(a^2-2(a')^2C^4),\quad (k_+ +k_-)^2=-(a')^2C^4,\quad k_+ +k_-=ia'C^2.
$$
Therefore
$$
D(\lam_2)=a(a+2a'C^2)+2i(a+a'C^2)(k_++k_-)-4k_+k_-=
$$
$$
a(a+2a'C^2)-2(a+a'C^2)a'C^2-a^2+2(a')^2C^4=0
$$
and then $\lam_2$ are real roots of $D(\lam)$.
Hence, the case $a'>a/C^2$ is {\bf linearly unstable}.
\\
2) Further consider the case $a'<-a/C^2$.
Then
$$k_+k_-=\frac 14(a^2-2(a')^2C^4)<0,\quad k_+ +k_-=-ia'C^2,$$
$$
D(\lam_2)=a^2+2aa'C^2+2(a+a'C^2)a'C^2-a^2+2(a')^2C^4
=4a'C^2(a+a'C^2)\not =0
$$
Therefore, in this case  $\lam_2$ are not roots of $D(\lam)$.
\begin{cor}
i)  In the unstable case $a'>a/C^2$:
both $\lam_2$ are roots of $D(\lam)$.
\\
ii) If $a'<-a/C^2$
then neither of the $\lam_2$  are roots of $D(\lam)$.
\end{cor}
Summarizing, we have proved the following result
\begin{theorem}\label{Spectr}
i)  If $a'\in (-\infty,a/(\sqrt 2C^2))$
 the only root of $D(\lam)$
is $\lam =0$ with multiplicity $2$.
\\
ii) If $a'\in [a/{\sqrt 2 C^2},a/C^2)$,
there are four roots of $D(\lam)$: zero (multiplicity
two) and $\pm i|\lam_2|$ (pure imaginary) with $\lam_2$ as in (\ref{lam}).
\\
iii) If $a'=+a/C^2$,
 the only root of $D(\lam)$
is $\lam =0$ multiplicity $4$.
\\
iv)
If $a'\in (a/C^2,+\infty)$, there are four
roots of $D(\lam)$: zero  (multiplicity
two) and $\pm|\lam_2|$ with $\lam_2$ as in (\ref{lam}).
In particular there exists a positive root (linear instability).
\end{theorem}

\begin{remark}
Imagine reducing $a'$ starting from a value greater than
$a/C^2$. Initially there are two real roots, $\pm|\lam_2|$, which
approach zero as $a'\to a/C^2$ from above, giving rise to an increase
of the multiplicity of the $\lam=0$ root to four when $a'=a/C^2$. As
$a'$ is reduced further below $a/C^2$ these two roots reappear as
a pair of conjugate pure imaginary roots which move from zero to
$\pm i\om$ as $a'$ goes from $a/C^2$ to $a/\sqrt 2C^2$. When
$a'=a/\sqrt 2C^2$ these two roots touch the branch point (end of the
continuous spectrum) and move onto an ``unphysical'' branch (on which
the conditions (\ref{re}) do not hold). As $a'$ is reduced further
these roots do not return to the physical branch and thus even when
their magnitude becomes zero they do not coalesce with the physical
$\lam=0$ root to increase its multiplicity and most importantly the
spectrum is pure continuous apart from zero for $a'<a/C^2$.
\end{remark}


\end{document}